\documentclass[fleqn,usenatbib]{mnras}
\usepackage{newtxtext,newtxmath}
\usepackage[T1]{fontenc}
\DeclareRobustCommand{\VAN}[3]{#2}
\let\VANthebibliography\thebibliography
\def\thebibliography{\DeclareRobustCommand{\VAN}[3]{##3}\VANthebibliography}

\usepackage[dvipdfmx]{graphicx}	
\usepackage{amsmath}	
\usepackage{color}
\usepackage{threeparttable}

\newcommand{\msun}{{\rm M_\odot}} 
\newcommand{\rsun}{{\rm R_\odot}} 
\newcommand{\yr}{{\rm yr}} 
\newcommand{\Myr}{{\rm Myr}} 
\newcommand{\au}{{\rm AU}} 
\newcommand{\kms}{{\rm km~s^{-1}}} 
\newcommand{\kelvin}{{\rm K}} 

\newcommand{\ms}{M_\ast}
\newcommand{\rs}{R_{\ast}}
\newcommand{\teff}{T_{\rm eff}}
\newcommand{\mdot}{\dot{M}_\ast}
\newcommand{\mdotc}{\dot{M}_{\rm crit}}

\newcommand{\rmin}{r_{\rm min}}
\newcommand{\rmax}{r_{\rm max}}
\newcommand{\rb}{r_{\rm B}}

\newcommand{\HI}{{\rm H}}
\newcommand{\HII}{{\rm H^+}}
\newcommand{\HeI}{{\rm He}}
\newcommand{\HeII}{{\rm He^+}}
\newcommand{\HeIII}{{\rm He^{++}}}
\newcommand{\ele}{{\rm e^-}}
\newcommand{\Hmi}{{\rm H^-}}
\newcommand{\Hmol}{{\rm H_2}}
\newcommand{\Hmolp}{{\rm H_2^+}}

\newcommand{\mbh}{M_\bullet}

\defcitealias{Wise2019Nature}{W19}
\defcitealias{Sakurai2020MNRAS}{SHI20}
\defcitealias{Toyouchi2020MNRAS}{Toyouchi et al. 2020}

\title[Radiative feedback on SMS formation]{Radiative feedback on supermassive star formation: \\
the massive end of the Population III initial mass function}

\author[D.~Toyouchi et al.]{
Daisuke~Toyouchi$^{1}$\thanks{E-mail: toyouchi@resceu.s.u-tokyo.ac.jp},
Kohei~Inayoshi$^{2}$,
Wenxiu~Li$^{2}$,
Zolt\'an~Haiman$^{3}$,
Rolf~Kuiper$^{4}$
\\
$^{1}$Research Center for the Early Universe (RESCEU), The University of Tokyo,
Hongo, 7-3-1, Bunkyo-ku
Tokyo, 113-0033, Japan\\
$^{2}$Kavli Institute for Astronomy and Astrophysics, Peking University, Beijing 100871, China\\
$^{3}$Department of Astronomy, Columbia University, New York, NY 10027, USA\\
$^{4}$Faculty of Physics, University of Duisburg-Essen, Lotharstra\ss e 1, D-47057 Duisburg, Germany
}

\date{Accepted XXX. Received YYY; in original form ZZZ}

\pubyear{2022}

\begin{document}
\label{firstpage}
\pagerange{\pageref{firstpage}--\pageref{lastpage}}
\maketitle

\begin{abstract}
Supermassive stars (SMSs) with masses of $\ms \simeq 10^4$--$10^5~\msun$ are invoked as possible seeds of high-redshift supermassive black holes, 
but it remains under debate whether their protostar indeed acquires sufficient mass via gas accretion overcoming radiative feedback.
We investigate protostellar growth in dynamically heated atomic-cooling haloes (ACHs) found in recent cosmological simulations,
performing three-dimensional radiation hydrodynamical (RHD) simulations that consider stellar evolution under variable mass accretion.
We find that one of the ACHs feeds the central protostar at rates exceeding a critical value, above which the star evolves in a cool bloating phase and hardly produces ionizing photons. 
Consequently, the stellar mass reaches $\ms \gtrsim 10^4~\msun$ unimpeded by radiative feedback.
In the other ACH, where the mass supply rate is lower, the star spends most of its life as a hot main-sequence star, emitting intense ionizing radiation.
Then, the stellar mass growth is terminated around $500~\msun$ by photoevaporation of the circumstellar disk.
A series of our RHD simulations provide a formula of the final stellar mass
determined either by stellar feedback or their lifetime
as a function of the mass supply rate from the parent cloud in the absence of stellar radiation.
Combining the results with the statistical properties of SMS-forming clouds in high-redshift quasar progenitor haloes,
we construct a top-heavy mass distribution of primordial stars over $\ms \simeq 100$--$10^5~\msun$, 
approximately following a power-law spectrum of $\propto \ms^{-1.3}$ with a steeper decline at $\ms \gtrsim 2 \times 10^4~\msun$.
Their massive BH remnants would be further fed via the dense debris disk, powering ``milli-quasars" with a bolometric luminosity of
$L_{\rm bol}~\gtrsim~10^{43}~{\rm erg~s^{-1}}$.
\end{abstract}




\section{Introduction}\label{sec:intro}

The initial mass function (IMF) of the first generation of stars, referred to as Population III (hereafter PopIII) stars, is the most fundamental information that determines the characteristics of the first galaxies and governs the evolution of the early universe.
The formation of PopIII stars is thought to take place in mini-haloes with virial masses of $M_{\rm vir} = 10^5$--$10^6~\msun$ 
at redshifts of $z~\sim~20$--$30$ \citep[e.g.,][]{Haiman1996ApJ, Abel2002Sci, Yoshida2003ApJ},
and their embryo protostars grow in mass quickly via accretion from the parent primordial gas clouds \citep[e.g.,][]{Omukai2003ApJ, Tan2004ApJ}.
The mass of the final product is determined by competition between the mass supply rate onto the protostar and 
the mass-loss rate from the circumstellar disk owing to photoheating by stellar irradiation \citep[e.g.,][]{Omukai2002MNRAS, McKee2008ApJ,
Hosokawa2011Sci, Stacy2012MNRAS, Fukushima2018MNRAS}.
According to recent theoretical studies, PopIII stars tend to be born with a typical mass around 
$\ms \sim 10$--$100~\msun$ \citep[e.g.,][]{Susa2014ApJ, Hirano2014ApJ, Hirano2015MNRAS}.
However, the PopIII-IMF shape at the high-mass end still remains a crucial question for seeding mechanisms of supermassive black holes
(SMBHs) with $\mbh \gtrsim 10^9~\msun$ observed in the brightest quasars (QSOs) at $z \gtrsim 6$ 
\citep[e.g.,][see also \citealt{Volonteri2012Sci, Haiman2013ASSL, Inayoshi2020ARA&A} for reviews from theoretical aspects]{Fan2001AJ, Willott2010AJ, Mortlock2011Natur, Wu2015Natur, Matsuoka2016ApJ, Banados2018Natur, Onoue2019ApJ, Wang2021ApJ, Yang2021ApJ}.

A possible mechanism of SMBH formation relies on the formation of ``heavy seed" formation in massive primordial haloes with 
$M_{\rm vir} \gtrsim 10^7~\msun$, where the gas is heated to $T \simeq 10^4~\kelvin$, enough to excite line transitions of atomic hydrogen,
the so-called atomic cooling haloes (ACHs).
When warm and massive gas in an ACH collapses owing to its self-gravity, the protostellar core is fed by the parent cloud at a rate of
$\mdot \simeq c_{\rm s}^3/G \simeq 0.1~\msun~\yr^{-1} ( T/10^4~\kelvin )^{3/2}$, 
where $c_{\rm s}$ is the sound speed of gas and $G$ is the gravitational constant,
which is sufficiently high to form supermassive stars (SMSs) with masses of $\ms \gtrsim 10^4~\msun$.
In the framework of modern galaxy formation, such massive objects are expected to form from primordial but molecular-hydrogen (H$_2$)-free gas clouds 
in protogalaxies since those compositions lead to efficient radiative cooling and thus induce vigorous cloud fragmentation. 
As H$_2$ dissociating processes, previous studies have extensively investigated the development of intense Lyman-Werner (LW) background radiation 
\citep[][]{Omukai2001ApJ, Oh2002ApJ, Bromm2003ApJ, Dijkstra2008MNRAS, Shang2010MNRAS, Wolcott-Green2011MNRAS,
Latif2013MNRAS, Regan2014ApJ, Sugimura2014MNRAS, Inayoshi2014MNRAS, Becerra2015MNRAS}.
In this scenario, ACH haloes forming SMSs are required to be irradiated with nearby star-forming galaxies but not be chemically polluted and tidally disrupted yet by those neighboring galaxies \citep[][]{Visbal2014MNRAS, Chon2016ApJ, Regan2017NatAs}.
However, the stringent requirements are hardly achieved in the early epoch of the universe,
so it is unclear whether SMS formation with the aid of external LW irradiation would produce a sufficient number of seed BHs.

In addition to the H$_2$ photo-dissociating process, dynamical disturbance on collapsing clouds associated with 
successive halo mergers leads to a delay of cloud collapse, and its heating effect counter-balances with H$_2$ cooling,
enabling SMS formation \citep[][]{Yoshida2003ApJ,Fernandez2014MNRAS}.
Three-dimensional (3D) cosmological hydrodynamical simulations by \citet{Wise2019Nature} (hereafter \citetalias{Wise2019Nature}) 
found that dynamical heating keeps the interior gas as warm as $T\simeq 10^4~{\rm K}$ before the onset of cloud collapse,
even though $\Hmol$ is not fully dissociated by external LW radiation \citep[see also][]{Regan2020MNRAS}.
Similarly, supersonic baryonic steaming motion relative to dark matter causes a delay in the onset of gas accumulation and star formation 
in less massive haloes \citep[][]{Tanaka2014MNRAS, Hirano2017Sci, Schauer2017MNRAS, Inayoshi2018MNRAS}.
Recently, \cite{Lupi2021MNRAS} and \cite{Li2021ApJ} explored BH seed formation in highly biased, 
overdense regions of the universe, conducting semi-analytical studies based on merger trees of DM haloes generated with a cosmological N-body simulation and the extended Press-Schechter formalism, respectively. 
They found that a combination of dynamical heating and LW irradiation considerably enhances the number of ACHs where SMSs potentially form.

Although gravitational collapse of such massive, H$_2$-free clouds potentially leads to heavy seed BHs 
through the formation of SMSs,
it still remains unclear whether the newly born protostars can be fed from the warm parent cloud 
at a sufficiently high accretion rate of $\mdot \simeq 0.1~\msun~\yr^{-1} ( T/10^4~\kelvin )^{3/2}$,
overcoming radiative feedback from the growing stars themselves.
Unlike ordinary PopIII star formation, the production of ionizing photons from a rapidly accreting protostar at 
rates {\it constantly} exceeding a critical value of $\mdotc \approx 0.04~\msun~\yr^{-1}$ is quenched because of expansion of the stellar surface
\citep[][]{Omukai2003ApJ, Hosokawa2013ApJ, Schleicher2013A&A, Haemmerle2018MNRAS}.
In fact, rapid inflows deposit heat onto the stellar envelope and prevent it from contracting via radiative energy loss on
a Kelvin-Helmholtz (KH) timescale, $\tau_{\rm KH} \sim 10^3$--$10^{4}~\yr$.
As a result, the protostar evolves into a supergiant phase with an inflated envelope of the size of $R_\ast \simeq 100~\au$ 
and effective temperature of $\teff \simeq 5000~\kelvin$.
However, if the accretion rate is {\it highly variable} and becomes lower than the critical value in a time duration longer than the KH timescale, 
the star begins to contract and increase the surface temperature of $\teff \sim 10^5~\kelvin$, which is the level of 
that in the zero-age main sequence (ZAMS) star that produces intense ionizing radiation \citep{Sakurai2015MNRAS}.
For instance, the central collapsing region in a dynamically heated ACH found in \citetalias{Wise2019Nature}
yields a modest accretion rate since the gas cools down to $T\simeq 10^3~\kelvin$ via $\Hmol$-line cooling at the center, where
external LW radiation is attenuated.
By the end of their simulations, the mass inflow rate is found to decrease toward the center and fall down to $\mdot < \mdotc$
at the innermost region (see the bottom right panel of their Figure~4).
It requires long-term and high-resolution simulations of the protostellar growth phase via disk accretion
to draw a robust conclusion for the fate of the central star and the remnant.

Recently, \citet*[][hereafter \citetalias{Sakurai2020MNRAS}]{Sakurai2020MNRAS} has explored the late phase of mass accretion onto 
a protostar in a dynamically heated ACH simulated by \citetalias{Wise2019Nature}, 
using one-dimensional (1D) radiation hydrodynamical (RHD) simulations that self-consistently take into account the stellar expansion and
contraction depending on the accretion history.
According to their simulations, while the ionizing radiation flux from the protostar heats the ambient gas and suppresses mass accretion
through the parent cloud,
the ionized regions shortly collapse owing to ram pressure exerted by the inflowing neutral gas. 
As a result of failed radiative feedback, the central star reaches $\sim 10^4~\msun$ within the lifetime even though the time-averaged 
mass accretion rate is lower than the critical value and thus the star grows following its ZAMS-like evolutionary track.
However, we claim that the spherically-symmetric assumption is no longer valid once an accretion disk forms around the protostar
and the effect of stellar radiative feedback begins to operate.
In fact, while the protostar is fed via a dense circumstellar disk due to the shielding effect of stellar radiation, 
photoheated gas on the disk surface produces outflows toward the bipolar directions and escapes from the gravitational influence radius of the star.
The mass-loss process from the disk decelerates the growth of the star and reduces the final stellar mass \citep[e.g.,][]{Hosokawa2011Sci, Hosokawa2012ApJ}.
Additionally, the accretion history of the protostar shows a great diversity when the circumstellar disk fragments into smaller clumps via gravitational instability,
and such non-axisymmetric structures play an essential role in transporting mass and angular momentum in the disk
\citep[e.g.,][]{Stacy2010MNRAS, Clark2011Sci, Greif2011ApJ, Greif2012MNRAS, Sugimura2020ApJ}.
For instance, the mass accretion rate becomes highly variable in the clumpy disk, showing short bursts followed by relatively long quiescent phases 
\citep[e.g.,][]{Smith2012MNRAS, Vorobyov2013ApJ, Sakurai2016bMNRAS, Matsukoba2021MNRAS}.
This effect would promote suppression of mass accretion onto a rapidly growing protostar when the star contracts and emits
intense ionizing radiation in an interval of clump migration causing an accretion burst at $\mdot \gtrsim \mdotc$.

In this paper, we reexamined SMS formation in the two ACHs of \citetalias{Wise2019Nature}, 
extending the 1D RHD simulations done by \citetalias{Sakurai2020MNRAS} into 3D RHD simulations that explicitly include the effects of photoevaporation and disk fragmentation.
Our 3D RHD simulations successfully track the stellar mass growth for $1~\Myr$, generally long enough to cover the entire life of an isolated massive star.
Moreover, we explore several cases with different initial conditions of self-gravitating parent clouds, rescaling the density structure of the gas in 
one of the ACHs studied by \citetalias{Wise2019Nature}.
This enables us to extend the parameter space and understand the final product in those primordial halos under various circumstances.
Combining our simulation results with the statistical properties of SMS-forming clouds in high-$z$ QSO progenitor haloes, 
we predict the IMF of massive primordial stars extending to the high-mass end of $\ms \simeq 10^3-10^5~\msun$.
Those SMSs are expected to leave BHs with similar masses after exhaustion of nuclear fuel \citep[e.g.,][]{Heger2003ApJ, Belczynski2010ApJ, Spera2015MNRAS}
or the onset of general relativistic instability \citep[e.g.,][]{Shibata2002ApJ, Umeda2016ApJ, Woods2017ApJ, Woods2020MNRAS}.

The rest of the paper is organized as follows.
In \S~\ref{sec:method}, we first describe the numerical method and setup of our 3D RHD simulations.
We show our main results based on the numerical simulations in \S~\ref{sec:results}, and provide an analytical argument to derive
the relation between the mass accretion rate and the final stellar mass in \S~\ref{sec:mass_growth}.
In \S~\ref{sec:discussion}, we discuss the mass distribution function of massive primordial stars with masses between $10^2\lesssim \ms/\msun \lesssim 10^5$, 
the subsequent mass growth of their remnant BHs, and some caveats of our simulations.
Finally, our conclusions are summarized in \S~\ref{sec:summary}.


\section{Simulation Method}\label{sec:method}

\subsection{Radiation hydrodynamical simulations}\label{sec:HDS}

We perform three-dimensional radiation hydrodynamical simulations to explore gas inflows around a protostar embedded in a self-gravitating gas cloud.
Our simulations are conducted with the hydrodynamical simulation code {\tt PLUTO} 4.1 \citep[][]{Mignone2007ApJS}, 
which has been extensively applied to study massive star formation, the evolution of protoplanetary disks, and gas accretion onto intermediate-mass BHs 
\citep[e.g.,][]{Kuiper2010ApJ, Hosokawa2016ApJ, Sugimura2017MNRAS, Sugimura2018MNRAS, Fukushima2018MNRAS, 
Nakatani2018aApJ, Nakatani2018bApJ, Toyouchi2019MNRAS, Toyouchi2020MNRAS, Toyouchi2021ApJ, Inayoshi2022aApJ}.

In this study, we adopt spherical coordinates ($r$, $\theta$, $\phi$) where a protostar is located at the origin. 
The basic equations of hydrodynamics that we solve are the following: the equation of continuity,
\begin{eqnarray}
\frac{\partial \rho}{\partial t} + \nabla \cdot (\rho \mbox{\boldmath $v$}) = 0,  
\label{eq:mass_cons}
\end{eqnarray}
and the equations of motion,
\begin{eqnarray}
\frac{\partial \rho v_r}{\partial t} + \nabla \cdot (\rho v_r \mbox{\boldmath $v$}) = - \frac{\partial P}{\partial r} + \rho \frac{v^2_\theta + v^2_\phi}{r} + \rho g_r \ ,  
\label{eq:mom_cons_r}
\end{eqnarray}
\begin{equation}
\begin{split}
\frac{\partial \rho v_\theta}{\partial t} + \nabla \cdot (\rho v_\theta \mbox{\boldmath $v$}) = &- \frac{1}{r}\frac{\partial P}{\partial \theta} - \rho \frac{v_\theta v_r}{r} \\
&+ \rho \frac{v^2_\phi~{\rm cot}~\theta}{r} + \rho g_\theta \ ,  
\label{eq:mom_cons_t}
\end{split}
\end{equation}
\begin{equation}
\begin{split}
\frac{\partial \rho v_\phi}{\partial t} + \nabla \cdot (\rho v_\phi \mbox{\boldmath $v$}) = &- \frac{1}{r~{\rm sin}~\theta}\frac{\partial P}{\partial \phi} - \rho \frac{v_\phi v_r}{r} \\
&- \rho \frac{v_\phi v_\theta~{\rm cot}~\theta}{r} + \rho g_\phi \ ,
\label{eq:mom_cons_p}
\end{split}
\end{equation}
%
where $\rho$ is the gas density, $\mbox{\boldmath $v$} = (v_r, v_\theta, v_\phi)$ is the velocity vector, $P$ is the gas pressure,
and $\mbox{\boldmath $g$}  = (g_r, g_\theta, g_\phi)$ is the total external force that includes the gravity of the central protostar
($\mbox{\boldmath $g$}_{\ast} =-G\ms/r^2 \mbox{\boldmath $e$}_{r}$) and the underlying dark matter (DM) halo 
($\mbox{\boldmath $g$}_{\rm DM}= -\nabla \Phi_{\rm DM}$), the self-gravity of gas ($\mbox{\boldmath $g$}_{\rm sg}= -\nabla \Phi_{\rm sg}$), 
and the outward radiative force $\mbox{\boldmath $g$}_{\rm rad}$ exerted on the gas through absorption and electron scattering of photons.
The gravitational potential of the gaseous mass is calculated by solving the Poisson equation
\begin{eqnarray}
\nabla^2 \Phi_{\rm sg} = 4 \pi G \rho,
\label{eq:sg}
\end{eqnarray}
\citep[e.g.,][]{Mignone2007ApJS, Mignone2012ApJS}.

We also solve the energy equation of
\begin{eqnarray}
\frac{\partial E}{\partial t} + \nabla \cdot (H \mbox{\boldmath $v$}) = \rho~\mbox{\boldmath $v$} \cdot \mbox{\boldmath $g$} + \rho~(\Gamma - \Lambda),  
\label{eq:ene_cons}
\end{eqnarray}
where $E$ is the total (internal and kinetic) energy density, $H$ is the enthalpy per unit volume, 
and $\Gamma$ and $\Lambda$ the specific heating and cooling rates in units of erg s$^{-1}$ g$^{-1}$.
Star formation in the circumstellar region and irradiation from those stars are not considered in this study, but their potential effect 
on the mass growth of the central protostar is discussed in Section \ref{sec:discussion}.
Additionally, we turn gas cooling off when the (local) Jeans length becomes unresolved with the largest size of each grid cell,
as implemented in the 3D RHD simulation of PopIII star formation \citep[][]{Hosokawa2016ApJ}.

The heating and cooling rates are estimated by solving a chemical reaction network of metal-free gas that is composed of the following nine species: 
$\HI$, $\HII$, $\HeI$, $\HeII$, $\HeIII$, $\ele$, $\Hmol$, $\Hmolp$, and $\Hmi$.
The number density of the $i$-th species $n_i$ is calculated with the non-equilibrium rate equation of
\begin{eqnarray}
\frac{\partial n_i}{\partial t} + \nabla \cdot (n_i \mbox{\boldmath $v$}) = n_{\rm H} R_i, 
\label{eq:chem_cons}
\end{eqnarray}
where $R_i$ is the sum of the reaction rate coefficients related to the $i$-th species and $n_{\rm H}$ is the number density of hydrogen nuclei.
We here consider the chemical reactions introduced as Nos. 1--32 in Table A1 of \citet{Glover2008MNRAS}, adopting the case B recombination rates for $\HII$, $\HeII$ and $\HeIII$.
We also take into account photoionization of $\HI$, $\HeI$, $\HeII$ and $\Hmol$, $\Hmi$ photo-detachment, and $\Hmol$ photodissociation, of which the absorption cross sections are summarized in Table 1 of \citetalias{Sakurai2020MNRAS}.
With the updated chemical abundances, we compute $\Lambda$ and $\Gamma$, including line-cooling by $\HI$, $\Hmol$, $\Hmolp$ and $\HeI$, recombination cooling of $\HII$, $\HeII$ and $\HeIII$, free-free emission, collisional ionization cooling of $\HI$, $\HeI$, $\HeII$ and $\Hmol$ dissociation cooling, $\Hmol$ formation heating, photoionization heating of $\HI$, $\HeI$, $\HeII$ and $\Hmol$, $\Hmi$ photo-detachment heating, $\Hmol$ photodissociation heating, and include self-shielding \citep[][]{Abel1997NewA, Omukai2000ApJ, Glover2007ApJ, Glover2008MNRAS}.

\subsection{Radiative feedback}\label{sec:RF}

Our simulation adopts a subgrid model to incorporate radiative feedback on the accretion flow.
We regard the interior of the innermost radius as a sink region, in which the central protostar and its circumstellar disk are contained.
The mass growth rate of the central protostar during each timestep, $\Delta t$, is simply evaluated from $\Delta \ms = \mdot \Delta t$, where $\mdot$ is the inward mass flux measured at the inner boundary.
Then, we consider radial propagation of photons emitted from the growing protostar and the unresolved accretion disk.
The stellar radiation flux at a photon frequency $\nu$ at the innermost radius $\rmin$ is given by
\begin{eqnarray}
F_{\ast, \nu} = \pi \left ( \frac{\rs}{\rmin} \right )^2 B_{\nu} (\teff), 
\label{eq:Fstellar}
\end{eqnarray}
where $\rs$ is the stellar radius, $\teff$ is the effective temperature, and $B_{\nu}$ is the Planck function.
The stellar radius and effective temperature are estimated with a stellar evolution model described in \S~\ref{sec:SE}.
We describe the disk radiation flux with a standard disk model, where the radiation spectrum is well approximated as
\begin{eqnarray}
F_{{\rm disk}, \nu} = \frac{1}{6\pi \rmin^2 [(\nu_\ast/\nu_{\rm min})^{4/3}-1]\nu_{\rm min}} \frac{G \ms \mdot}{\rs} \left(\frac{\nu}{\nu_{\rm min}}\right)^{1/3}, \nonumber \\
(\nu_{\rm min} \leq \nu \leq \nu_\ast)
\label{eq:Fdisk}
\end{eqnarray}
where $\nu_{\rm min}$ is the lowest frequency we consider and the cutoff frequency is given by
\begin{eqnarray}
\nu_\ast &=& 3.14 \times 10^{15}~{\rm Hz} \nonumber \\
&\times& \left(\frac{\ms}{1~\msun}\right)^{1/4}
\left(\frac{\mdot}{10^{-1}~\msun~\yr^{-1}}\right)^{1/4}
\left(\frac{\rs}{1~\rsun}\right)^{-3/4}
\label{eq:nucrit}
\end{eqnarray}
\citep[e.g.,][]{Shakura1973A&A, Kato2008bhad.book}.
We note that the cutoff frequency is as low as $\nu_{\ast}\simeq 10~{\rm eV}/h$ throughout our simulations, where $h$ is the Planck constant.
Therefore, the emergent disk radiation plays only a minor role in photoionization and heating of the surrounding gas.
The total flux injected from the innermost radius is set to $F_{\rm in, \nu} = F_{\ast, \nu} + F_{\rm disk, \nu}$.

We solve the multi-frequency radiative transfer equations, considering a frequency range of $h\nu_{\rm min}(=0.04~{\rm eV}) \leq h\nu \leq h\nu_{\rm max}(=118~{\rm eV})$.
The number of frequency bins is $N_\nu = 50$. 
We designed an non-uniform frequency grid layout: the frequency bin size is finer near the ionization threshold energy of H (13.6 eV), He (24.6 eV), and He$^+$ (54.4 eV).
In the frequency ranges of $h\nu < 11.2~{\rm eV}$ and $13.6~{\rm eV}<h\nu $, the radiation flux density can analytically be expressed as
\begin{eqnarray}
F_\nu = \left ( \frac{\rmin}{r} \right )^2 F_{\rm in, \nu} ~ \exp \left [ - \sum_i N_{i} \sigma_{i, \nu} \right ], 
\label{eq:Flux}
\end{eqnarray}
where $\sigma_{i, \nu}$ and $N_{i}$ are the absorption cross section of the $i$-th species and its column density evaluated from 
the innermost radius $\rmin$ to a radius $r$. 
We ignore electron scattering in the radiative transfer calculation since the cross section of electron scattering, $\sigma_{\rm es} = 6.65 \times 10^{-25}~{\rm cm^2}$, is much smaller than that of absorption by neutral hydrogen even at $\nu = \nu_{\rm max}$, $\sigma_{{\rm H}, \nu} \sim 10^{-20}~{\rm cm^{2}}$.
We note that the component of diffusive EUV radiation ($h\nu > 13.6~{\rm eV}$) produced by radiative recombination of gas 
is not considered because diffusive photons are typically negligible compared to direct ones emitted from the central protostar.
The radiation flux absorbed by each component is used for calculating the photoionization rates, photodetachment rates owing to 
lower energy photons, and photoheating rate ($\Gamma_{\rm ph}$), respectively.

In the LW frequency band of $11.2$--$13.6~{\rm eV}$, we consider a representative frequency of H$_2$ dissociating photons 
($h\nu_{\rm LW}=12.4$ eV) instead of solving multifrequency radiative transfer.
The radiation flux is calculated by
\begin{eqnarray}
F_{\nu_{\rm LW}} = \left ( \frac{\rmin}{r} \right )^2 F_{\rm in, \nu_{\rm LW}} f_{\rm sh},
\label{eq:Flux_LW}
\end{eqnarray}
where $f_{\rm sh}$ characterizes the effect of H$_2$ self-shielding and H shielding,
and the approximated function form is given by
\begin{eqnarray}
f_{\rm sh} &=& f_{\rm sh, H_2} \cdot f_{\rm sh, H}, \\[5pt]
f_{\rm sh, H_2} &=& \frac{0.965}{(1+x_{\rm H2}/b_5)^{1.1}} \nonumber \\
&+& \frac{0.035}{(1+x_{\rm H2})^{0.5}}~{\rm exp} \left [-8.5 \times 10^{-4}(1 + x_{\rm H2})^{0.5} \right ], \\[5pt]
f_{\rm sh, H} &=& (1 + x_{\rm H})^{-1.6}~{\rm exp} \left (-0.15~x_{\rm H} \right), 
\label{eq:fshield}
\end{eqnarray}
where $x_{\rm H2} \equiv N_{\rm H_2} / (5 \times 10^{14}~{\rm cm^{-2}})$, $x_{\rm H} \equiv N_{\rm H}/(2.85 \times 10^{23}~{\rm cm^{-2}})$, 
$b_5 \equiv \sqrt{kT/m_{\rm p}}/(10^5~{\rm cm~s^{-1}})$, $k$ is the Boltzmann constant, and $m_{\rm p}$ is the proton mass
\citep[][]{Wolcott-Green2019MNRAS}.
The H$_2$ photodissociation and heating rates through the two-step Solomon process are calculated by 
\begin{eqnarray}
k_{\rm pd} = 1.1 \times 10^8~\frac{F_{\nu_{\rm LW}}}{\rm erg~s^{-1}~cm^{-2}~Hz^{-1}}~{\rm s^{-1}}, 
\label{eq:photodis_rate}
\end{eqnarray}
and 
\begin{eqnarray}
\Gamma_{\rm pd} = 6.4 \times 10^{-13}~n_{\rm H_2}~k_{\rm pd}~{\rm erg~s^{-1}~cm^{-3}},
\label{eq:photodis_heat}
\end{eqnarray}
\citep[][]{Abel1997NewA}.

The outward radiation pressure force along the radial direction is given by,
\begin{eqnarray}
g_{\rm rad, r} = \frac{n_{e}}{c}\int \sigma_{\rm es} F_{\nu} {\rm d}\nu + \frac{\Gamma_{\rm ph}}{c}.
\label{eq:rad_force}
\end{eqnarray}
The tangential components of the radiative force are neglected in this study.

\subsection{Stellar evolution}\label{sec:SE}

\begin{table*}
\begin{center}
\caption{
Stellar radii and effective temperatures for ZAMS stars with different stellar mass. 
The data is taken from \citet{Marigo2001A&A} for $\ms \leq 100~\msun$ and \citet{Bromm2001ApJ} for $\ms = 300$ and $1000~\msun$.
At $\ms \geq 10^4~\msun$, the values of $\rs$ and $\teff$ are set by linearly extrapolating in the logarithmic quantities.
}
\label{table:zams}
\begin{tabular}{ccccccccccc} \hline \hline
$\ms~(\msun)$ & 2 & 10 & 30 & 50 & 100 & 300 & 1000 & $10^4$ & $10^5$ & $10^6$ \\ \hline
${\rm log}~\rs~(\rsun)$ & $-8.93\times 10^{-4}$ & 0.139 & 0.323 & 0.451 & 0.627 & 0.959 & 1.20 & 1.66 & 2.13 & 2.59 \\
${\rm log}~\teff~(\kelvin)$ & 4.14 & 4.65 & 4.87 & 4.93 & 4.98 & 5.05 & 5.07 & 5.11 & 5.15 & 5.19 \\
\hline \hline \\
\end{tabular}
\end{center}
\end{table*}

\begin{table*}
\begin{center}
\caption{
Stellar radii and effective temperatures for a supergiant protostar that grows at a constant mass accretion rate of $\mdot = 0.1~\msun~\yr^{-1}$
(the data taken from \citealt{Hosokawa2013ApJ}). 
At $\ms \geq 10^5~\msun$, the values of $\rs$ and $\teff$ are set by linearly extrapolating in the logarithmic quantities.
}
\label{table:rgb}
\begin{tabular}{ccccccccc} \hline \hline
$\ms~(\msun)$ & 2 & 10 & 20 & 27 & 100 & $1.7\times10^4$ & $10^5$ & $10^6$ \\ \hline
${\rm log}~\rs~(\rsun)$ & 2.30 & 2.26 & 2.37 & 2.78 & 3.32 & 4.34 & 4.69 & 5.15\\
${\rm log}~\teff~(\kelvin)$ & 3.65 & 3.69 & 3.70 & 3.68 & 3.68 & 3.80 & 3.84 & 3.90 \\
\hline \hline \\
\end{tabular}
\end{center}
\end{table*}

\begin{table*}
\begin{center}
\caption{
Time-averaged mass accretion rates and stellar masses obtained from the 1D HD and 3D RHD simulations,
where $\langle \cdot \rangle$ means the time-averaged value over $0\leq t \leq t_{\rm avg}$.
We adopt $t_{\rm avg} = 0.5~\Myr$ for most cases except the LWH-10 model setting $t_{\rm avg} = 0.2~\Myr$, which corresponds to the termination time of the 3D RHD simulation.
The tabulated values of stellar mass is calculated by $\ms = \langle \mdot \rangle \times 0.5~\Myr$. 
}
\label{table:model}
\begin{tabular}{ccccc} \hline \hline
Models & 
$\left < \mdot \right >_{\rm 1HD}$ & 
$M_{\rm \ast, 1HD}$ & 
$\left < \mdot \right >_{\rm 3RHD}$ & 
$M_{\rm \ast, 3RHD}$ \\
& $[\msun~\yr^{-1}]$ & $[\msun]$ & $[\msun~\yr^{-1}]$ & $[\msun]$ \\ \hline
MMH & $7.7\times10^{-2}$ & $3.8\times10^{4}$ & $3.1\times10^{-2}$ & $1.6\times10^{4}$ \\
LWH & $2.2\times10^{-2}$ & $1.1\times10^{4}$ & $1.0\times10^{-3}$ & $5.1\times10^{2}$ \\
LWH-0.1 & $2.7\times10^{-4}$ & $1.3\times10^{2}$ & $3.3\times10^{-5}$ & $17$ \\
LWH-0.5 & $5.6\times10^{-3}$ & $2.8\times10^{3}$ & $1.8\times10^{-4}$ & $89$ \\
LWH-2 & $6.1\times10^{-2}$ & $3.1\times10^{4}$ & $1.8\times10^{-2}$ & $9.0\times10^{3}$ \\
LWH-10 & $0.56$ & $2.8\times10^{5}$ & $0.19$ & $9.5\times10^{4}$ \\
\hline \hline \\
\end{tabular}
\end{center}
\end{table*}

\begin{figure}
\begin{center}
\includegraphics[width=\columnwidth]{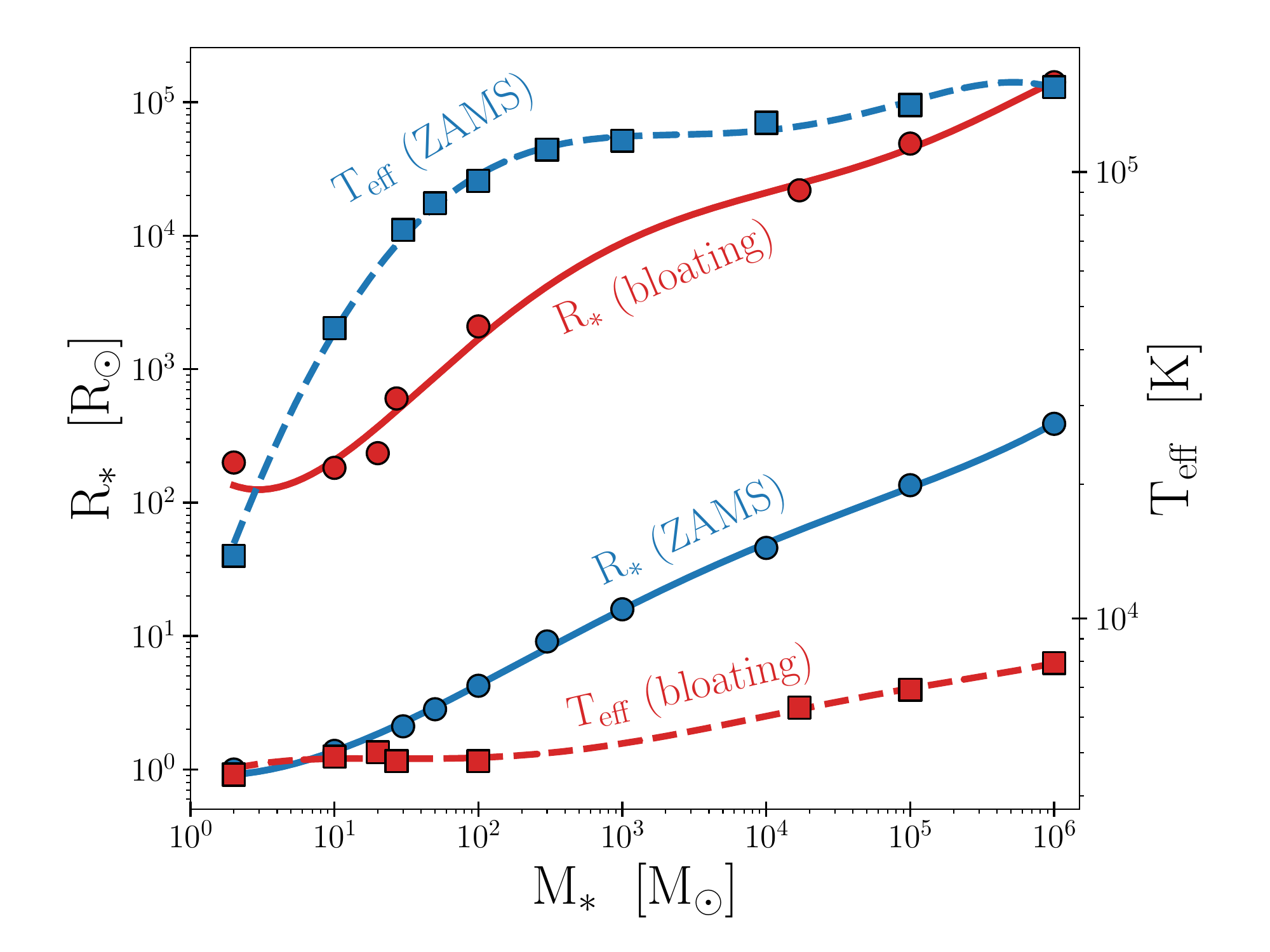}
\end{center}
\vspace{-5mm}
\caption{
The fourth-degree polynomial fitting functions that characterize the stellar evolution for the ZAMS (blue) and supergiant phases (red).
The solid and dashed curves represent the results of the stellar radius ($R_\ast$) and effective temperature ($T_{\rm eff}$), respectively (the data taken from 
Tables~\ref{table:zams} and \ref{table:rgb}).
}
\label{fig:R_T}
\end{figure}

The protostellar radius and effective temperature are updated at each timestep, depending on the mass accretion history.
When the accretion rate is lower than $\mdotc~(\equiv 0.04~\msun~\yr^{-1})$, the stellar structure is thermally relaxed and 
their properties are assumed to be those in a zero-age main sequence (ZAMS) phase.
On the other hand, when the accretion rate is higher than the critical value, the stellar surface is bloated owing to rapid entropy injection
through the accreting matter, and thus the surface temperature becomes as low as $T_{\rm eff} \simeq 5000~{\rm K}$ 
\citep[][]{Hosokawa2013ApJ}.
In Tables~\ref{table:zams} and \ref{table:rgb}, we summarize the data of $R_\ast$ and $T_{\rm eff}$ for various masses 
for the ZAMS phase and the bloating supergiant phase accreting at $\mdot=0.1~\msun~\yr^{-1}$, respectively.
Throughout this paper, the protostellar evolution in each phase is described by fitting the data with 
fourth-degree polynomial functions as shown in Figure~\ref{fig:R_T}.

In our study, we adopt either of the two radii for a given stellar mass, depending on whether the accretion rate exceeds $\mdotc$.
For instance, when the accretion rate onto a protostar abruptly rises due to clump migration through a disk, we assume that the star 
expands instantaneously and its size reaches that in the super-giant phase.
This assumption is reasonable because stellar expansion occurs on the accretion timescale in the surface layer, 
$\tau_{\rm acc} \sim 10^2$--$10^3~\yr$ \citep[e.g.,][]{Hosokawa2013ApJ, Sakurai2015MNRAS}, which is 
as short as the dynamical timescale at the innermost radius in our simulations.
On the other hand, when the accretion rate drops below $\mdotc$ and a super-giant star evolves its ZAMS phase,
we take into account the KH contraction phase on a finite timescale as
%
\begin{eqnarray}
\frac{\rs}{\rsun} = \left ( \frac{1}{R_{\ast, 0}/\rsun} + \frac{t-t_0}{\tau_{\rm c}} \frac{1}{M_{\ast, 0}/\msun} \right )^{-1},
\label{eq:Rs_evol}
\end{eqnarray}
where $R_{\ast, 0}$ and $M_{\ast, 0}$ are the stellar radius and mass in the super-giant phase before the contraction 
and $\tau_{\rm c} = 3.3 \times 10^{2}~\yr$ is the typical KH timescale in that phase \citep[][]{Sakurai2015MNRAS}.
The minimum stellar radius is set to that in the ZAMS phase with the corresponding mass.
Note that the stellar mass is almost constant during quiescent phases and thus $M_{\ast, 0}=M_{\ast}$ is imposed to derive Eq.~(\ref{eq:Rs_evol}).
Following the size evolution, the surface temperature is calculated by logarithmic interpolation (i.e., $\log R_\ast - \log T_{\rm eff}$) between the two phases.

Based on hydrodynamical simulations, a hydrostatic protostar with $\simeq 0.2~\msun$ is first formed at the center of a massive
atomically-cooling gas cloud but the protostar grows to $\sim 2~\msun$ in a few years \citep{Inayoshi2014MNRAS, Becerra2015MNRAS}.
We assume that the initial stellar mass is $\ms = 2~\msun$ and the stellar structure follows that in its ZAMS phase.

\subsection{Grid configuration and Initial conditions}\label{sec:initial}

We set a computational domain of $\rmin \leq r \leq \rmax$, $0 \leq \theta \leq \pi/2$, and $0 \leq \phi \leq 2 \pi$.
In the radial direction and polar direction, we set up logarithmically spaced grids to achieve a high resolution 
near the center and equator. In the azimuthal direction, we adopt uniformly spaced grids.
The number of grid cells in each direction is set to $(N_r, N_\theta, N_\phi) = (200, 36, 72)$, enabling us to resolve 
the disk thickness with at least three grid cells.
As our fiducial case, we set $\rmin = 10^3~{\rm AU}$ and $\rmax = 10^7~{\rm AU}$.
We have checked the convergence of our simulation results, varying the number of the grid cells 
and the size of the innermost radius (see Appendix~\ref{sec:cc}).
We adopt the innermost radius so that the stellar Bondi radius for cold gas with a temperature of $T=200$ K 
is sufficiently resolved in the initial stage ($M_\star \simeq 2~\msun$);
%
\begin{eqnarray}
\rb = \frac{G \ms}{c^2_{\rm s}} 
\sim 2.5 \times 10^3~{\rm AU} \left ( \frac{\ms}{2~\msun} \right ) \left ( \frac{T}{200~\kelvin} \right )^{-1} \ ,
\label{eq:rb}
\end{eqnarray}
where we assume isothermal gas with a polytropic index $\gamma = 1$ and the mean molecular weight of $\mu = 2.3$, 
yielding the sound speed $c_{\rm s} = \sqrt{\gamma k_{\rm B} T / (\mu m_{\rm p})} = 0.85 (T/200~\kelvin)^{1/2}~{\rm km \ s^{-1}}$.

We adopt outflow boundary conditions at the innermost and outermost cells, where zero gradients across the boundaries are imposed on physical quantities to allow gas to flow out from the computational domain.
On the other hand, gas inflows through the boundaries are prohibited by imposing $v_r \leq 0$ and $v_r \geq 0$ at the inner and outer boundaries, respectively.
On the pole and equator, i.e., $\theta = 0$ and $\pi/2$, we impose reflective conditions.
We also adopt periodic boundary conditions for the azimuthal direction at $\phi = 0$ and $2\pi$.

We mainly explore protostellar evolution in two primordial haloes, the more irradiated halo (LWH) and the more massive halo (MMH), originally identified in \citetalias{Wise2019Nature}.
The initial conditions for our 3D RHD simulations are taken from the spherically-averaged gas cloud profiles of the LWH and MMH model,
as shown in Figure~4 of \citetalias{Wise2019Nature}.
We fit the profiles of the enclosed gas mass (${\rm log}~r$ - ${\rm log}~M_{\rm enc}$), radial velocity (${\rm log}~M_{\rm enc}$ - $v_r$), 
temperature (${\rm log}~M_{\rm enc}$ - ${\rm log}~T$), and $\Hmol$ fraction (${\rm log}~M_{\rm enc}$ - ${\rm log}~X_\Hmol$) with seventh-degree polynomial functions.
We assume that the initial ionization degree is approximated as $X_\ele \simeq \left (2 \alpha_{\rm rec} n_\HI t_{\rm ff} \right )^{-1}$ 
in a collapsing cloud \citep[][]{Inayoshi2011MNRAS}, where $\alpha_{\rm rec}$ is the recombination rate coefficient, 
and $t_{\rm ff}$ is the free-fall time of the cloud.
Since the density distribution follows a self-similar profile of $n_\HI \propto r^{-2}$ (and thus $t_{\rm ff} \propto r$), 
we set the initial profile of the ionization degree to $X_\ele = 10^{-7}~(r/\rmin)$ and impose $X_{\HII}=X_\ele$ because of charge neutrality.
The initial number density relative to hydrogen nuclei is set to 0.0833, and helium is assumed to be initially all neutral.

\begin{figure}
\begin{center}
\includegraphics[width=\columnwidth]{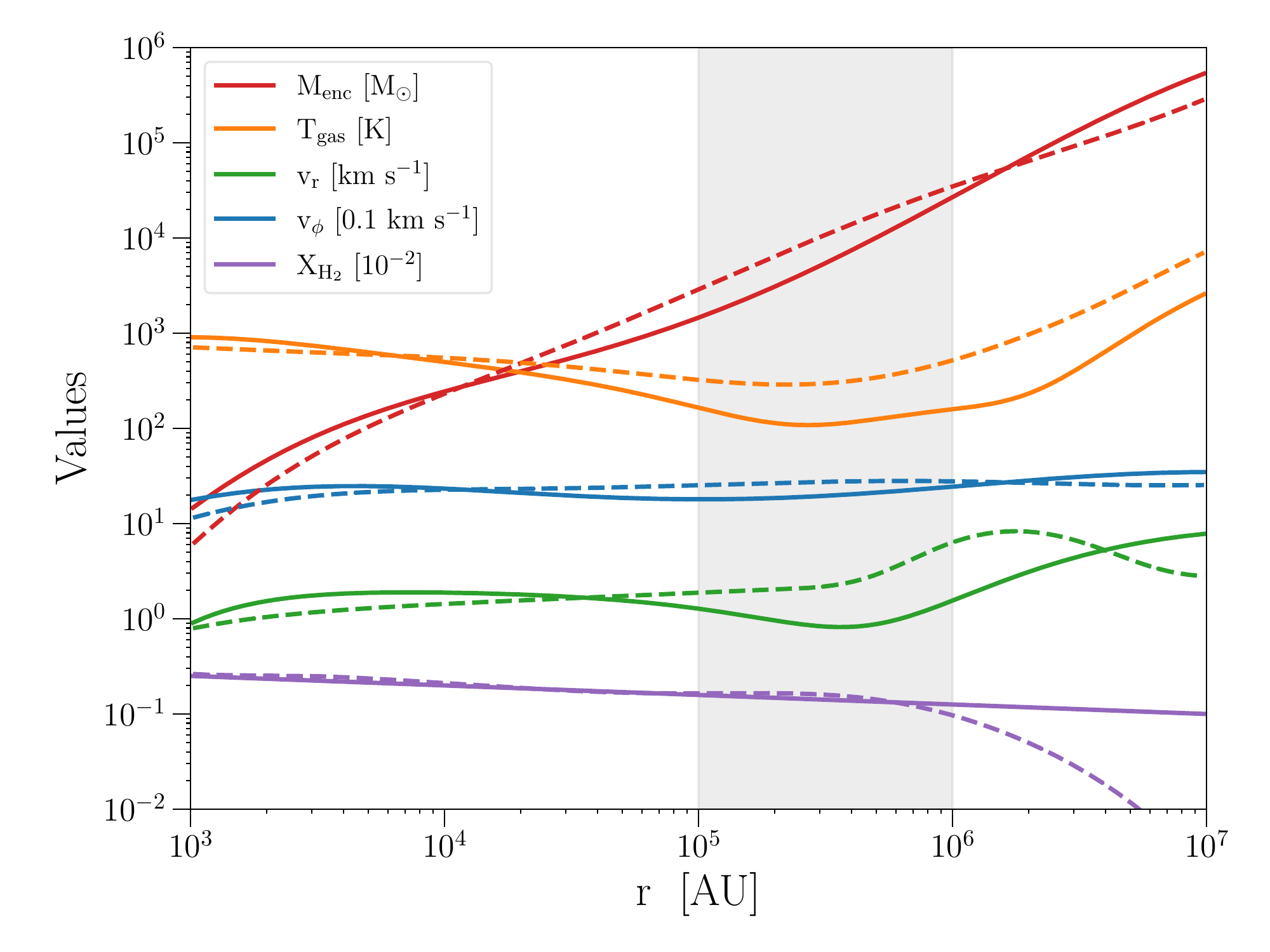}
\end{center}
\vspace{-5mm}
\caption{
The initial conditions used for the LWH (solid) and MMH (dashed) models.
Each colored curve represents the radial profiles of different physical quantities; 
the gas enclosed mass $M_{\rm enc}$ (red), temperature $T$ (orange), radial velocity $v_r$ (green), cylindrical rotational velocity $v_\phi$ (blue), 
and $\Hmol$ fraction $X_{\Hmol}$ (purple).
Note that $v_\phi$ and $X_{\Hmol}$ are normalized by $0.1~{\rm km~s}^{-1}$ and 0.01, respectively, for visualization purpose.
The grey shaded area indicates the range over $r = 10^5$--$10^6~\au$, where one free-fall timescale of $\simeq 0.1$--$1~\Myr$ is comparable to 
the stellar lifetime.
}
\label{fig:IC}
\end{figure}

For rotational motion of the clouds, we do not refer to that shown in \citetalias{Wise2019Nature}, 
since our spherically-averaged initial condition cannot take into account the rotationally supported disk already appearing within $r \sim 10^4~\au$ at the last snapshot of the original simulation.
Instead, we suppose a cylindrical rotation of the clouds, of which the angular momentum vector is aligned with the $Z$-axis at the initial condition.
The rotational velocity $v_\phi$ at any $R = r {\rm sin} \theta$ is set to half of the Keplerian velocity evaluated on the equatorial plane ($\theta = \pi/2$), 
i.e., $v_\phi(R) = 0.5 \sqrt{G M_{\rm enc}(r = R)/R}$, which is universally realised in gravitational gas collapse in primordial haloes 
\citep[][]{Abel2002Sci, Yoshida2008Sci}.
The initial profile of each physical quantity assumed in LWH and MMH is shown in Figure~\ref{fig:IC}.

The two target haloes experience several episodes of rapid mass assembly, and thus the onset of gravitational collapse is delayed until
 $z = 15.3~(16.4)$, when the virial mass reaches $5.8~(2.6) \times 10^7~\msun$ for the LWH (MMH) halo.
In this paper, we incorporate the gravity of the DM halo with a Navarro-Frenk-White (NFW) potential \citep[][]{Navarro1997ApJ, Mo1998MNRAS}, 
\begin{eqnarray}
\Phi_{\rm DM}(r) =  - \frac{2 k_{\rm B} T_{\rm vir}}{\mu m_{\rm p}}~
\frac{{\rm ln}(1+r/r_{\rm s})}{r/r_{\rm s}}~f(c) \ ,
\label{eq:phi_dm}
\end{eqnarray}
where $T_{\rm vir}$ and $r_{\rm vir}$ are the virial temperature and radius of the halo, 
$r_{\rm s} (\equiv r_{\rm vir}/c)$ is the characteristic radius of the NFW density profile,
$c$ is the concentration parameter, and 
$f(c) \equiv c/[{\rm ln}(1+c) - c/(1+c)]$.
We model the dependence of the mean concentration parameter on virial mass $M_{\rm vir}$ and redshift $z$: 
$c \simeq 1.56~(M_{\rm vir}/10^9~\msun)^{-0.13}~[(1+z)/21]^{-1}$
\citep[][]{Bullock2001MNRAS}.
We note that for both target haloes, the gaseous self-gravity dominates the gravity of the DM halo within $r \sim 10^6~\au$ and 
the free-fall time scale at the exterior is longer than our computational time of $1~\Myr$.
Therefore, the external gravitational force cased by the DM halo plays a minor role in determining gas dynamics.

We do not consider an external LW radiation background in our simulation, though the LWH and MMH haloes are exposed to
modest LW irradiation of $J_{\rm LW} \lesssim 4\times 10^{-21}~{\rm erg~s^{-1}~cm^{-2}~Hz^{-1}~str^{-1}}$.
The level of LW intensity is not high enough to impact the chemo-dynamics of the collapsing cloud in the halo.
We also neglect turbulent motion in the cloud at the initial condition, though gas motion is highly turbulent in 
the original simulations by \citetalias{Wise2019Nature}.
In fact, a recent study by \citet{Regan2020bOJAp} explored protostar formation in the LWH and MMH haloes
and found that turbulent motion in the collapsing cloud affects its long-term evolution at $t \gtrsim 1~\Myr$.
Therefore, our simulations focus on the early evolutionary stage of $t \leq 1~\Myr$.

\subsection{Model cases}

In order to study stellar mass growth mechanisms under various environments, in addition to the LWH and MMH haloes, we also explore four other cases, where the initial conditions are generated by altering that of the LWH halo.
The models named \mbox{LWH-0.1}, \mbox{LWH-0.5}, \mbox{LWH-2}, and \mbox{LWH-10} employ the enclosed gas and DM mass at any radii scaled up or down by a factor of 0.1, 0.5, 2, and 10, respectively, from the original values.
In these synthesized models, rotational velocities are also altered according to the modified Keplerian velocity profiles 
while we adopt the same conditions as in the original LWH model for other quantities.

In addition to the 3D RHD simulations, we also perform 1D HD simulations to highlight the coupling effects of feedback from stellar radiation and disk formation due to angular momentum of accreting gas.
The initial conditions for those 1D calculations are identical to their 3D counterparts, except for neglecting rotational motion in the collapsing clouds.
For both the LWH and MMH haloes, we have tracked the evolution of the central star up to $t = 1~\Myr$, corresponding to the typical lifetime of massive stars,
while the other calculations have been terminated earlier to save computational cost.
Table~\ref{table:model} shows the time-averaged mass accretion rate onto and the mass of the central star obtained in our 1D HD and 3D RHD simulations for each model.
The values tabulated here are evaluated at $t = 0.5~\Myr$, except for the LWH-10 model, which has been terminated at $t = 0.2~\Myr$
since the numerical time step becomes too short due to the strong gravity of the density-enhanced cloud.
For this model, we present the stellar mass at $t = 0.5~\Myr$, estimated by assuming a constant mass growth in the time-averaged accretion rate evaluated at the the termination time, i.e., $\ms = \langle \mdot \rangle \times 0.5~\Myr$.

Here, we briefly comment the results of the 1D HD simulations for the LWH and MMH haloes.
For both target haloes, the central star grows to $\ms \gtrsim 10^4~\msun$ by $t = 0.5~\Myr$.
Note here that the MMH halo gives potentially more rapid mass accretion onto the central protostar than the LWH halo. 
This is because the MMH gas cloud has relatively high enclosed gas mass, radial infall velocity, and thus high inward mass flux at $r = 10^5$--$10^6~\au$, where the free-fall time is $\tau_{\rm ff} \equiv \sqrt{r^3/{\rm G} M_{\rm enc}} \sim 0.1$--$1~\Myr$ (see Figure~\ref{fig:IC}). 
The hydrodynamics at the radial scale governs the mass supply to the cloud center on the corresponding time scale.
As a result, the time-averaged accretion rates for the LWH and MMH haloes are $\langle \mdot \rangle_{\rm 1HD}~\sim~0.02$ and $0.08~\msun~\yr^{-1}$, being below and above $\mdotc$, respectively.
The difference in the potential mass supply rates from the parent clouds can render the fates of the central protostars much different between the two target haloes.
In the next section, we thoroughly investigate the accretion dynamics around the central star and its mass growth history in the LWH and MMH haloes with our 3D RHD simulations.

Finally, we also perform three runs with the same numerical setup as the LWH halo but adopting different grid configurations to check the numerical convergence of our simulations.
These numerical experiments show that our simulation results are sufficiently converged with less than 20~\% difference in the central star's mass at least until $t \sim 0.2~\Myr$ (see Appendix~\ref{sec:cc}).


\begin{figure*}
\begin{center}
\includegraphics[width=1.7\columnwidth]{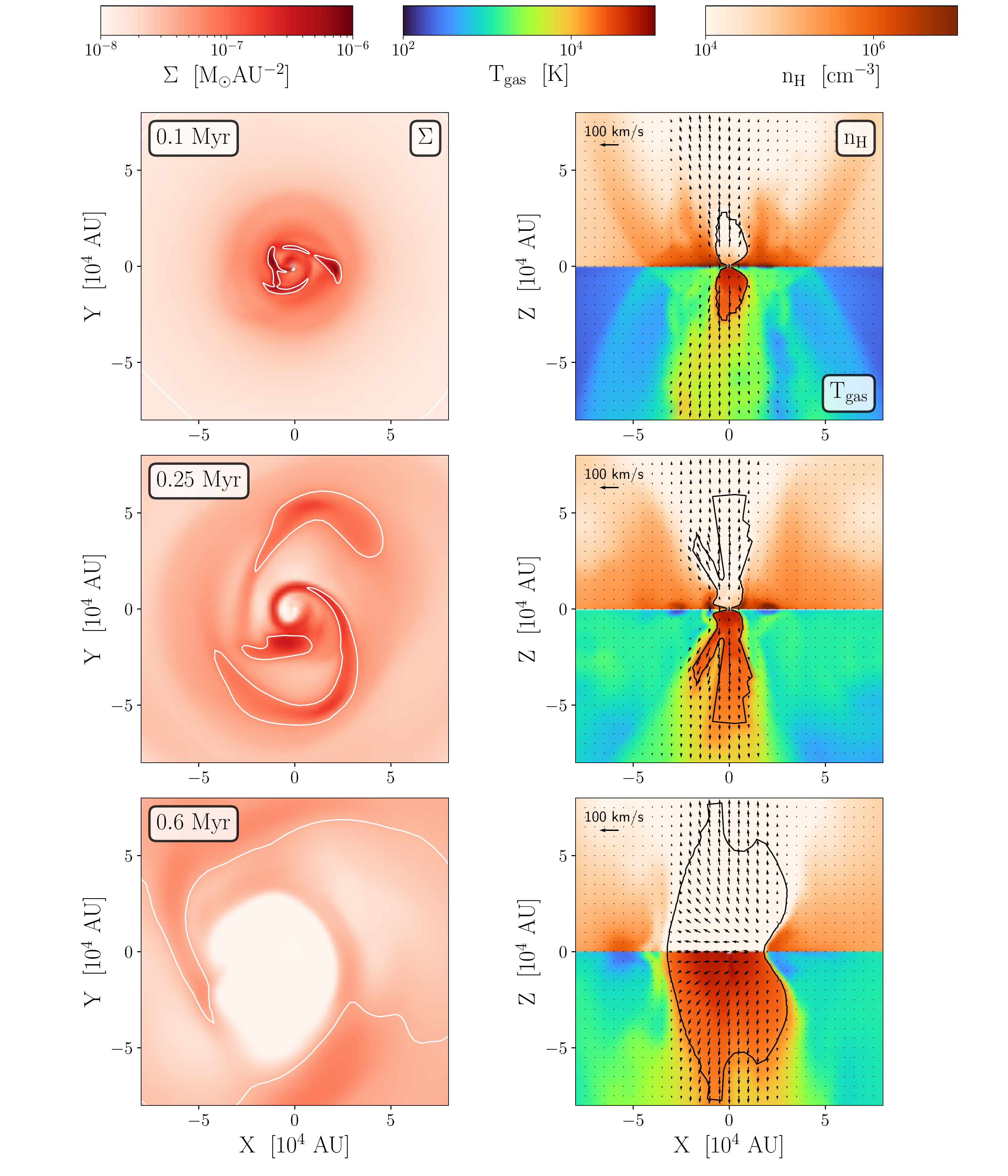}
\end{center}
\vspace{-5mm}
\caption{
Time evolution of the density and thermal structure in the LWH halo: from the top to the bottom panels, the elapsed time is $t =$ 0.1, 0.25, and 0.6 Myr. 
The left panels show the face-on distribution of the surface mass density, and 
the right panels represent the edge-on distribution of the number density (top) and temperature (bottom), sliced at $Y=0$.
The white contours in the left panels indicate the locally unstable regions, where the Toomre-$Q$ value is below unity.
In the right panels, we overlay the velocity vectors with a reference speed of $100~\kms$ and the location of the ionization front where
the electron fraction is 0.99 (biconical regions denoted by the black contour).
As the central star grows and emits more intense ionizing radiation, the circumstellar disk suffers mass loss owing to photoevaporation.
As a result, the inner part of the disk is evacuated by the time of $t = 0.6~\Myr$.
}
\label{fig:str_LWH}
\end{figure*}

\begin{figure}
\begin{center}
\includegraphics[width=\columnwidth]{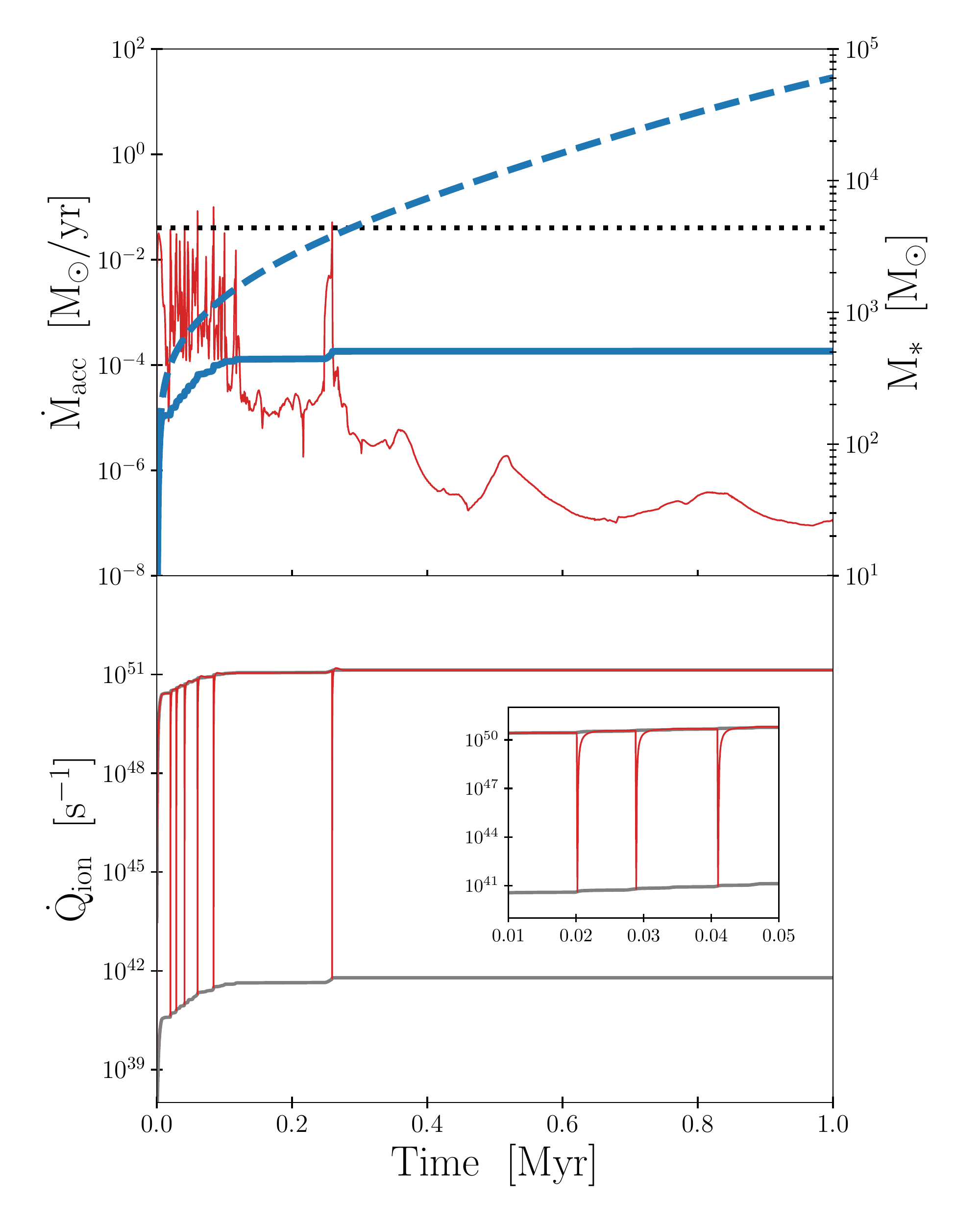}
\end{center}
\vspace{-5mm}
\caption{
{\it Top panel}: 
Time evolution of the mass accretion rate onto the central star (red solid) and the stellar mass (blue solid) obtained from the 3D RHD simulation for the LWH halo.
The black dotted line indicates the critical mass accretion rate for stellar expansion, $\mdotc = 0.04~\msun~\yr^{-1}$, and 
the blue dashed curve represents the stellar mass obtained from the 1D simulation without radiative feedback for the same halo.
{\it Bottom panel}: 
Time evolution of the production rate of ionizing photon from the accreting star (red).
The upper and lower grey curves represent the values of $\dot{Q}_{\rm ion}$ for the ZAMS and supergiant phases, respectively.
The inset zooms in to show the short-time evolution of $\dot{Q}_{\rm ion}$ at $0.01\leq t/\Myr \leq 0.05$, where multiple stellar expansion and contraction episodes occur owing to intermittent accretion bursts exceeding the critical rate.
}
\label{fig:acc_LWH}
\end{figure}

\begin{figure}
\begin{center}
\includegraphics[width=\columnwidth]{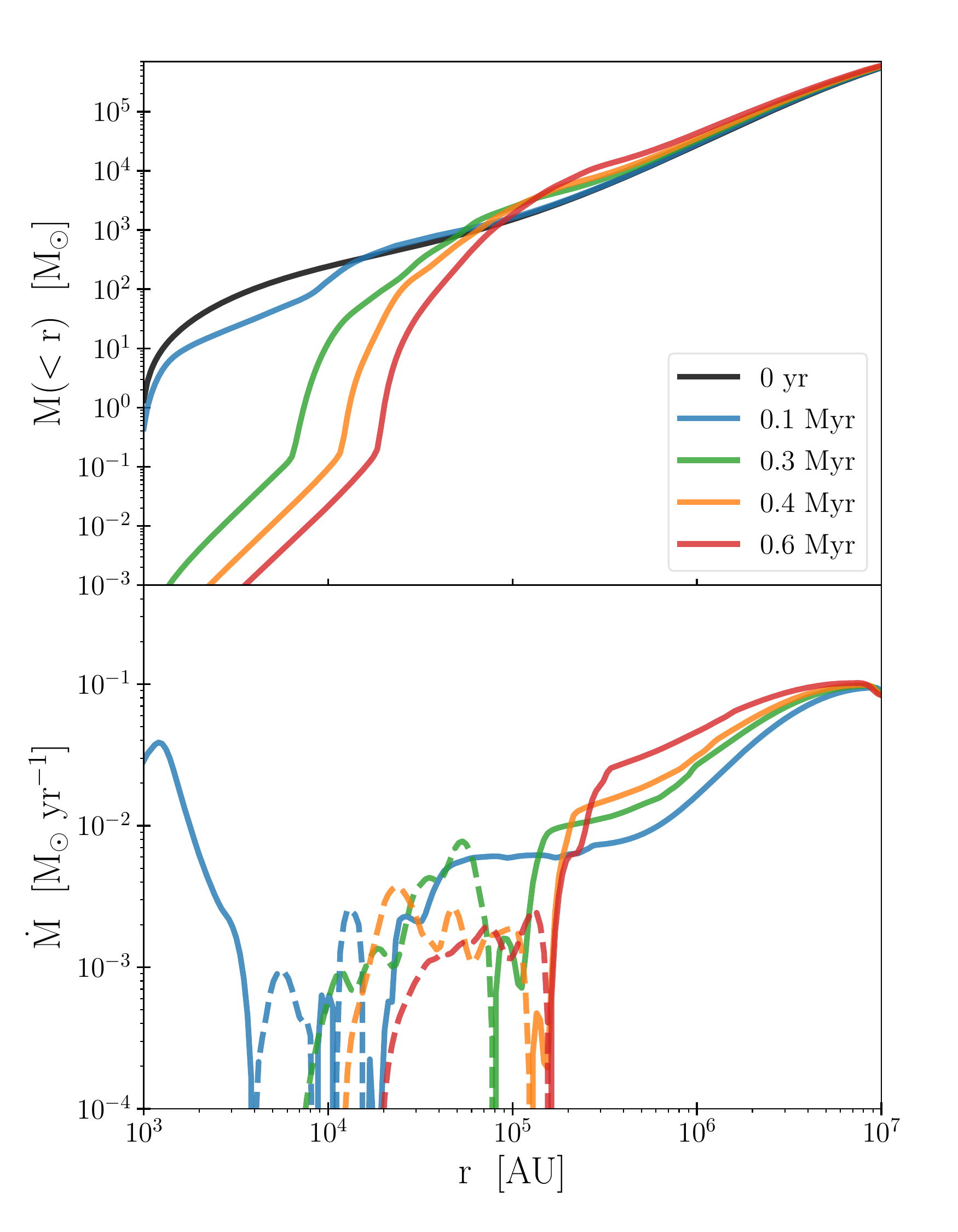}
\end{center}
\vspace{-5mm}
\caption{
Radial profiles of the spherical enclosed mass (top) and radial mass flux (bottom) in the LWH halo
at five different elapsed times.
The solid and dashed curves in the bottom panel indicate the inflow and outflow rates, respectively.
The central region is evacuated by the end of the simulation, when the ionization bubble reaches $r\simeq 2\times 10^5~{\rm AU}$
and the mass supply from the radius is suppressed.
}
\label{fig:cm_LWH}
\end{figure}

\section{Results}\label{sec:results}

\subsection{LWH halo}\label{sec:LWH}

Figure~\ref{fig:str_LWH} shows the density and temperature distribution of the accretion flows in the central region of the LWH halo 
at three elapsed times, of which the face-on and edge-on views are exhibited in the left and right columns, respectively.
The angular momentum of the collapsing cloud leads to the formation of an accretion disk around the central protostar.
In the early stage ($t = 0.1~\Myr$; top panels), the circumstellar disk becomes dense enough to be gravitationally unstable 
and thus gravitational torque caused by forming spiral waves with a spatial extent of $10^4~\au$ drives mass transport toward the central star.
As the protostar grows, the emergent ionizing radiation heats the surrounding matter and produces bipolar outflows of hot gas.
At $t = 0.25~\Myr$ (middle panels), the ionized bubbles further expand toward the polar directions and reach $|Z| \gtrsim 5 \times 10^4 ~\au$.
Once the ionization front breaks the stellar gravitational influence radius of $r_{\rm B} \sim 10^4~\au$ for ionized gas, the heated gas begins to evaporate 
from the disk surface and the disk mass decreases.
As shown in the face-on view, a low density cavity forms around the star owning to mass loss caused by the strong pressure-gradient 
force of ionized gas.
By the late stage ($t = 0.6~\Myr$; bottom panels), the inner-most edge of the accretion disk gradually moves outward and the central region within 
$\simeq 5 \times 10^4~{\rm AU}$ is fully evacuated.
Note that the ionized gas shows a coherent flow pattern along the X-axis near the equator.
This is caused by the pressure gradient owing to density contrast at the edges of the ionized cavity (see the bottom panels).

\begin{figure*}
\begin{center}
\includegraphics[width=1.7\columnwidth]{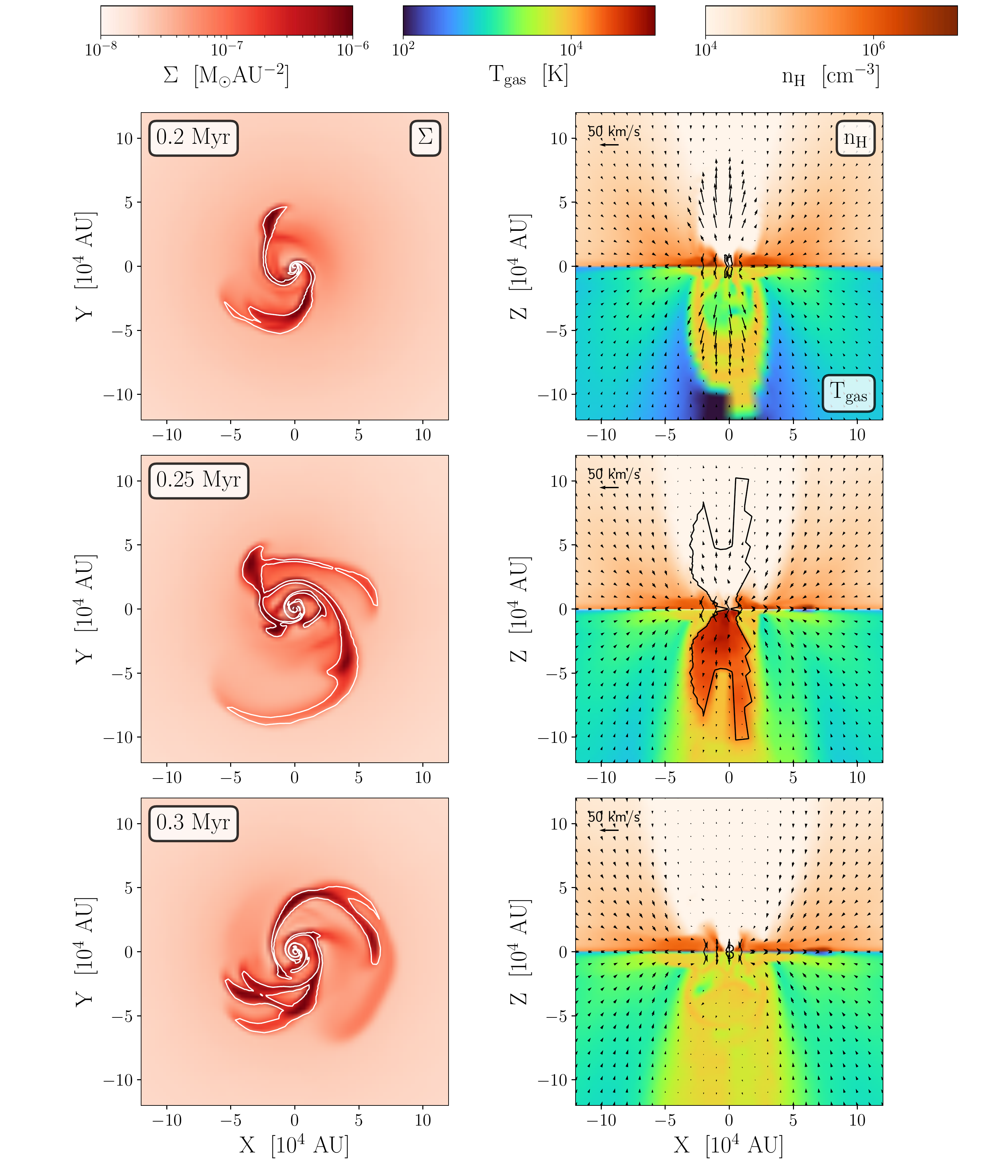}
\end{center}
\vspace{-5mm}
\caption{
Same as Figure~\ref{fig:str_LWH}, but for the results in the MMH halo:
from the top to the bottom panels, the elapsed time is $t =$ 0.2, 0.25, and 0.3~Myr.
In contrast to the LWH halo, a dense accretion disk with prominent spiral wave structures
feeds the central protostar over a long period.
Transient production of ionized bubbles and bipolar outflows in a short time interval
is found but does not prevent the protostar from growing in mass significantly.
}
\label{fig:str_MMH}
\end{figure*}

The upper panel of Figure~\ref{fig:acc_LWH} shows the time evolution of the mass accretion rate onto the central star (red solid) 
and its mass (blue solid) obtained in the 3D RHD simulation.
For comparison, we overlay $\mdotc = 0.04~\msun~\yr^{-1}$ (black dotted) and the stellar mass growth in the 1D HD simulation (blue dashed).
In the early stage, the central star undergoes efficient mass growth since dense accretion flows reach the vicinity of the central star 
(see the top panels of Figure~\ref{fig:str_LWH}).
The accretion rate intermittently increases with a typical interval of $\sim 10^4~\yr$, which corresponds to the dynamical timescale at 
$R \sim 10^4~\au$ where spiral waves form, and occasionally exceed the critical accretion rate of $\mdotc$ to alter the stellar evolution.
By $t\simeq 0.1~\Myr$, the star grows to be as massive as $\ms \simeq 300~\msun$.
However, in the later phase, stellar mass growth is essentially quenched by stellar radiative feedback.
In fact, after the last accretion burst associated with the migration of a massive fragment at $t \sim 0.25~\Myr$, 
the accretion rate slumps below $\mdot \sim 10^{-6}~\msun~\yr^{-1}$ and never revives.
By the end of the simulation, the stellar mass reaches $\ms \simeq 500\msun$.

The lower panel of Figure~\ref{fig:acc_LWH} shows the production rate of ionizing photons by the central star.
During most of the epochs, the production rate of ionizing photons is as high as $\dot{Q}_{\rm ion} \sim 10^{51}~{\rm s^{-1}}$,
corresponding to that for a massive ZAMS star with $T_{\rm eff}\simeq 10^5~\kelvin$.
This high value reflects that the central accreting star is thermally relaxed and contracts to its ZAMS radius.
In contrast, when the mass accretion rate abruptly rises and exceeds the critical value of $\mdotc$ in a short time, 
the stellar radius is inflated so that the stellar effective temperature is as low as $T_{\rm eff}\simeq 5000~{\rm K}$ and 
thus the ionizing photon production rate sharply declines.
However, the ionizing photon production rate recovers to the ZAMS value in a KH time of $\tau_{\rm KH} < 10^4~\yr$ 
(see the inserted panel in Figure~\ref{fig:acc_LWH}),
which is typically shorter than the time interval between accretion bursts owing to disk fragmentation.
Therefore, in the LWH halo case, nearly constant production of ionizing photons from the massive protostar 
hinders its growth via accretion through the disk.

It is worth noting that the 1D simulation without stellar radiation produces an extremely massive star with 
$\ms \sim 6 \times 10^4~\msun$.
This result clearly shows that stellar mass growth is strongly suppressed by radiative feedback
when the time-averaged mass accretion rate in the early stage is lower than the critical rate for stellar evolution.
Moreover, even with stellar radiation in the 1D spherically symmetric simulation, the dynamics of the inflowing gas in the halo is hardly affected by 
radiative feedback because the photoionized gas is completely confined within the stellar Bondi radius
and collapses to the center owing to ram pressure of the incoming neutral gas (\citetalias{Sakurai2020MNRAS}). 

While the bipolar outflows driven by photoionization evacuate the central region, the ionization front does not reach the virial radius of the halo.
To show the entire structure of the surrounding gas, Figure~\ref{fig:cm_LWH} presents the radial profiles of 
the spherically-averaged enclosed mass (top) and mass flux (bottom) at five different elapsed times.
The ionized bubble propagates outward quickly in the earlier stage and its expansion ceases at $t \gtrsim 0.4~\Myr$.
By the end of the simulation, the bubble size nearly converges to $r \sim 2 \times 10^5~\au$,
which is $\simeq 20$ times larger than the gravitational influence radius of the central star for ionized gas 
but is sufficiently smaller than the size of the collapsing parent cloud.
Outside the ionized bubble, an overdense shell-like structure forms owing to deceleration of the outflow (dashed) and continuous inflowing gas (solid). 
We find that the shell structure stably exists within $t = 1~\Myr$ (i.e., during the stellar lifetime) but its mass increases
due to inflow of neutral gas from the exterior of the ionized bubble.
When the mass shell becomes sufficiently massive or the properties of incident radiation from the central star change 
(e.g., due to stellar collapse to a BH), runaway collapse of the massive gas cloud is triggered and thus the central object is fed
at a high rate, as reported in previous RHD simulations that study BH accretion under extremely dense environments 
\citep[e.g.,][]{Inayoshi2016MNRAS, Inayoshi2022aApJ, Toyouchi2019MNRAS, Toyouchi2020MNRAS}.
Based on this idea, we briefly discuss the subsequent growth of a BH left after the end of the central star's life in \S~\ref{sec:BHs}.

\begin{figure}
\begin{center}
\includegraphics[width=\columnwidth]{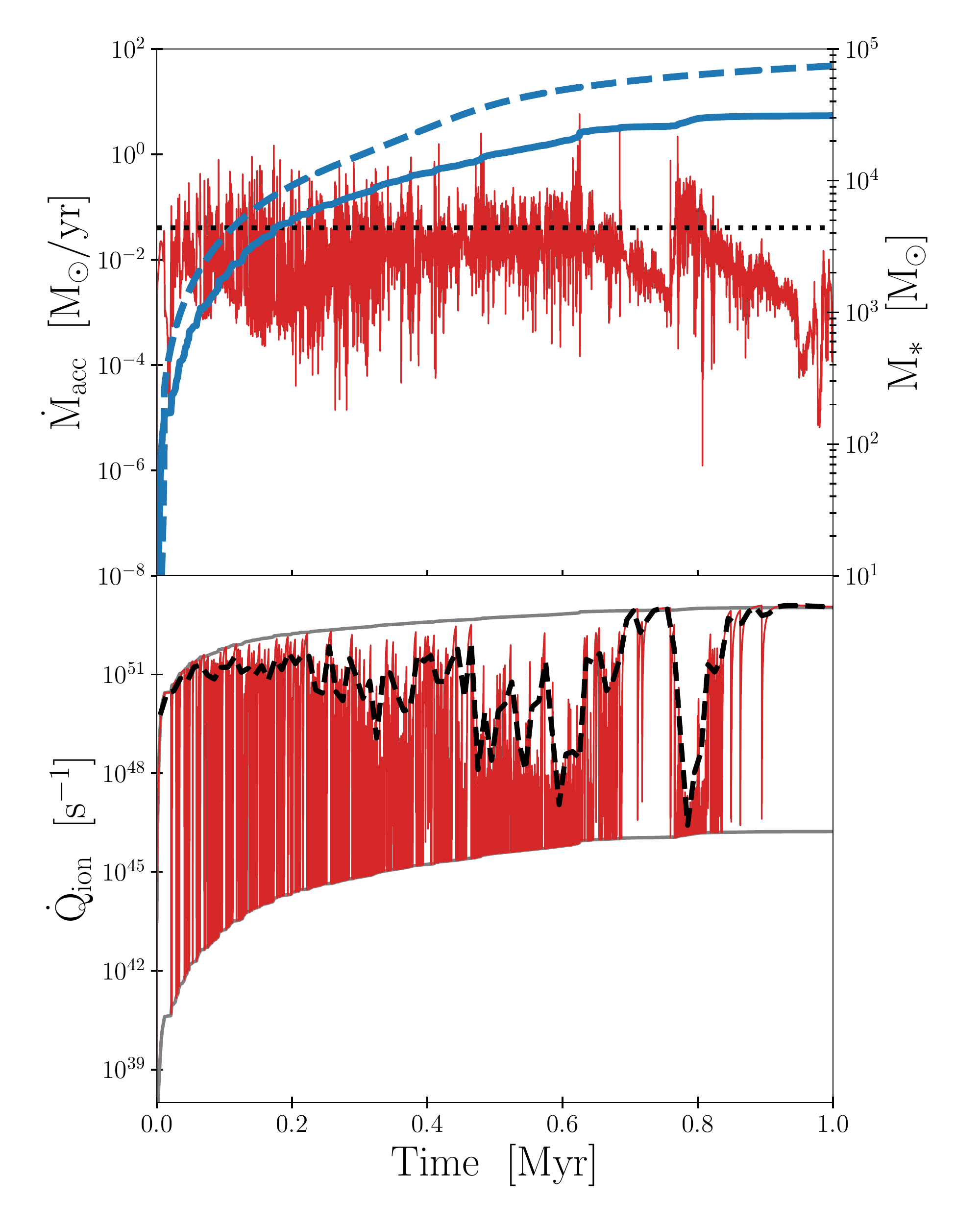}
\end{center}
\vspace{-5mm}
\caption{
Same as Figure~\ref{fig:acc_LWH}, but for the MMH halo.
The black dashed curve in the bottom panel shows the ionizing photon production rates averaged in 
each duration of $\Delta t = 10^4~\yr$.
For the MMH halo, rapid accretion episodes make the central star in its supergiant phase with a low effective temperature
of $T_{\rm eff}\simeq 5000~{\rm K}$, resulting in suppression of the emergent ionizing radiation flux to
$\dot{Q}_{\rm ion} < 10^{51}~\rm s^{-1}$.
}
\label{fig:acc_MMH}
\end{figure}

\begin{figure}
\begin{center}
\includegraphics[width=\columnwidth]{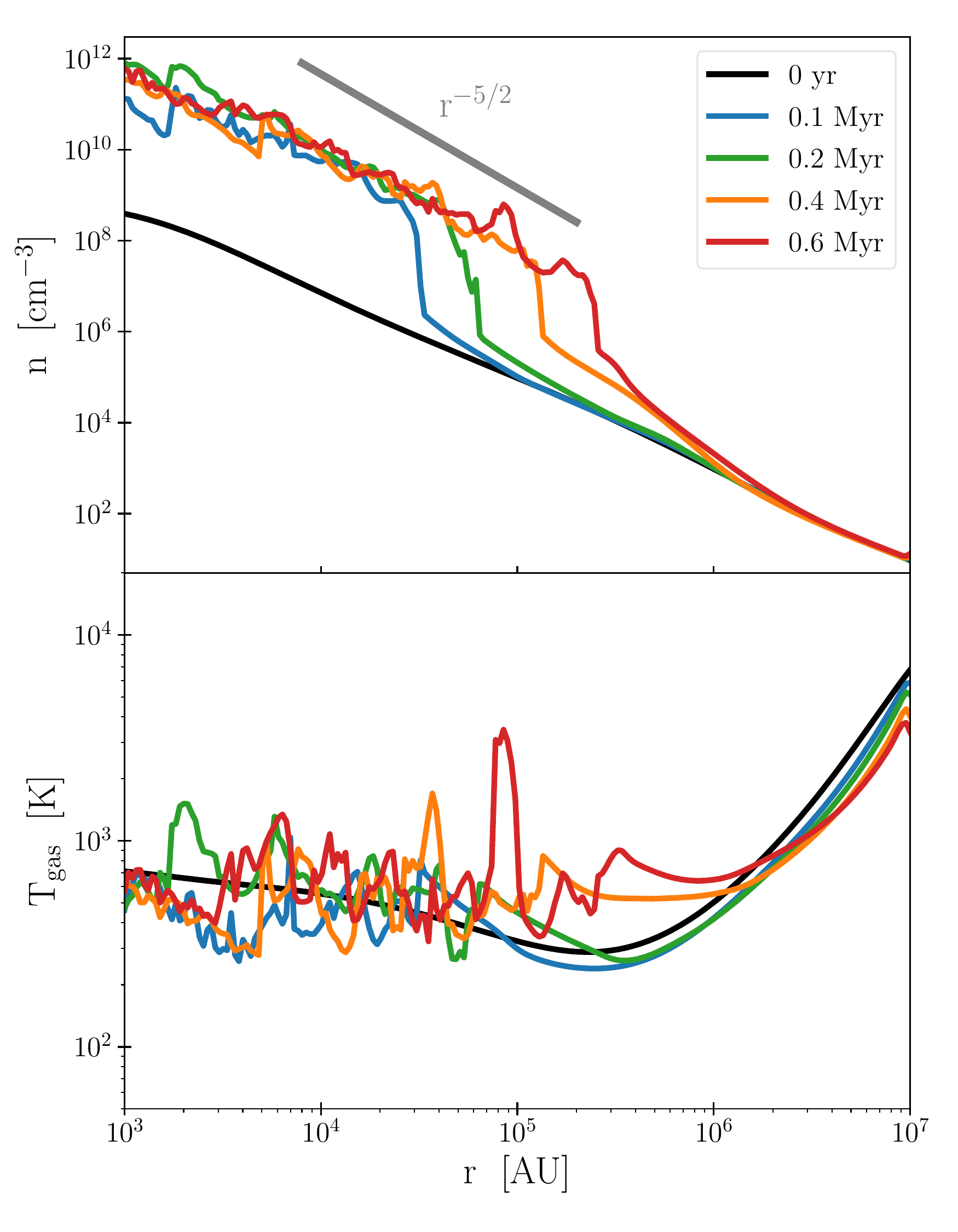}
\end{center}
\vspace{-5mm}
\caption{
Radial profiles of the azimuthal-averaged gas density (top) and temperature (bottom) along the equator (i.e., $\theta = \pi/2$) in the MMH halo
at five different elapsed times. 
The grey line in the top panel indicates a slope of $r^{-5/2}$, characterizing the density profile of a gravitationally unstable disk in a quasi-steady state (see \S~\ref{sec:MMH} for detail).
}
\label{fig:den_MMH}
\end{figure}

\subsection{MMH halo}\label{sec:MMH}

Next, we move on to the case in the MMH halo, where the mass inflow rate at larger radii is higher than in the LWH halo.
Figure~\ref{fig:str_MMH} presents the density and temperature distribution of the accretion flows in the MMH halo.
The surface density distribution (left panel) shows that prominent spiral waves form in a dense disk owing to 
gravitational instability even down to the inner region of $R < 10^4~\au$.
The non-axisymmetric structure transports angular momentum of the gas outward and thus brings the mass  
to the vicinity of the central protostar.
As the protostar grows in mass, low-density cavities are created in the polar directions and the maximum velocity 
of the outflow barely exceeds the sound speed.
However, ionized bubbles do not expand continuously but appear in an episodic manner, unlike in the LWH case.
Therefore, mass loss due to photoevaporation plays a minor role in the mass budget of the circumstellar disk.

Figure~\ref{fig:acc_MMH} shows the time evolution of stellar mass growth (top) and stellar ionizing flux (bottom) in the MMH halo.
An episodic behavior of the accretion rate is found in the early stage as in the LWH case.
In the MMH halo, however, owing to continuous mass supply through the dense disk and weak mass loss,
the burst-like accretion proceeds without being suppressed.
As a result, the time-averaged accretion rate becomes as high as $\langle  \mdot \rangle \simeq 0.03~\msun~\yr^{-1}$,
and the central star eventually grows to $\ms \simeq 3\times10^4~\msun$ by the end of the simulation at $t = 1~\Myr$.
The final mass is comparable to that obtained from the 1D simulation without stellar radiation ($\ms \sim 7\times10^4~\msun$; dashed curve),
suggesting that the effect of gas angular momentum (and partially radiative feedback) reduces the mass by a factor of $2-3$.
This result is a natural outcome of the long-term accretion bursts with a high frequency where the peak rates exceed the critical rate for stellar evolution. 
As shown in the bottom panel of Figure~\ref{fig:acc_MMH}, the ionizing photon production rate varies between the values for 
the ZAMS ($\mdot < \mdotc$) and the bloating super-giant phases ($\mdot \geq \mdotc$).
Due to the episodic behavior, the value of $\dot{Q}_{\rm ion}$ averaged in a time interval of $\Delta t = 10^4~\yr$ (black dashed) 
is comparable or lower than $10^{51}~{\rm s^{-1}}$ most of the time except in the last stage.
We note that this time interval is $\sim 10$ times longer than the KH timescale in super-giant phases,
so that the ionizing photon flux reaches the value for the ZAMS unless multiple accretion bursts at rates of $\mdot \geq \mdotc$
occur in shorter time durations of $\lesssim 10^4~\yr$.
For instance, the average value of $\dot{Q}_{\rm ion}$ drops significantly from that for the ZAMS phase 
at $t \sim 0.6~\Myr$ and $0.8~\Myr$ due to successive burst accretion.
Therefore, suppression of radiative feedback from the accreting protostar with a bloated and low-temperature surface 
enables the formation of an SMS in the MMH halo.

Owing to weak radiative feedback, continuous mass accretion develops a distinct disk structure in the MMH case.
To see the properties of the disk in a quasi-steady state, 
Figure~\ref{fig:den_MMH} presents the radial profiles of the density (top) and temperature (middle)
along the equator (i.e., $\theta = \pi/2$) at five different epochs.
Starting at the beginning of the simulation, the infalling gas with angular momentum accumulates around the central star 
and the outer edge of the disk moves outward because gas with higher angular momentum accretes onto the disk in a later phase.
The density structure can be well approximated by $\rho \propto r^{-5/2}$ within the disk (grey line) and
is a characteristic profile for a gravitationally unstable disk (see more details below).
In the quasi-steady state, the gas temperature is kept as cold as $T_{\rm gas} \sim 500~\kelvin$ within the disk,
where the gas becomes fully molecular ($x_{\rm H_2}\simeq 0.5$) in spiral structures; 
otherwise the molecular fraction is as low as $x_{\rm H_2}\simeq 10^{-4}-10^{-3}$.

\begin{figure}
\begin{center}
\includegraphics[width=\columnwidth]{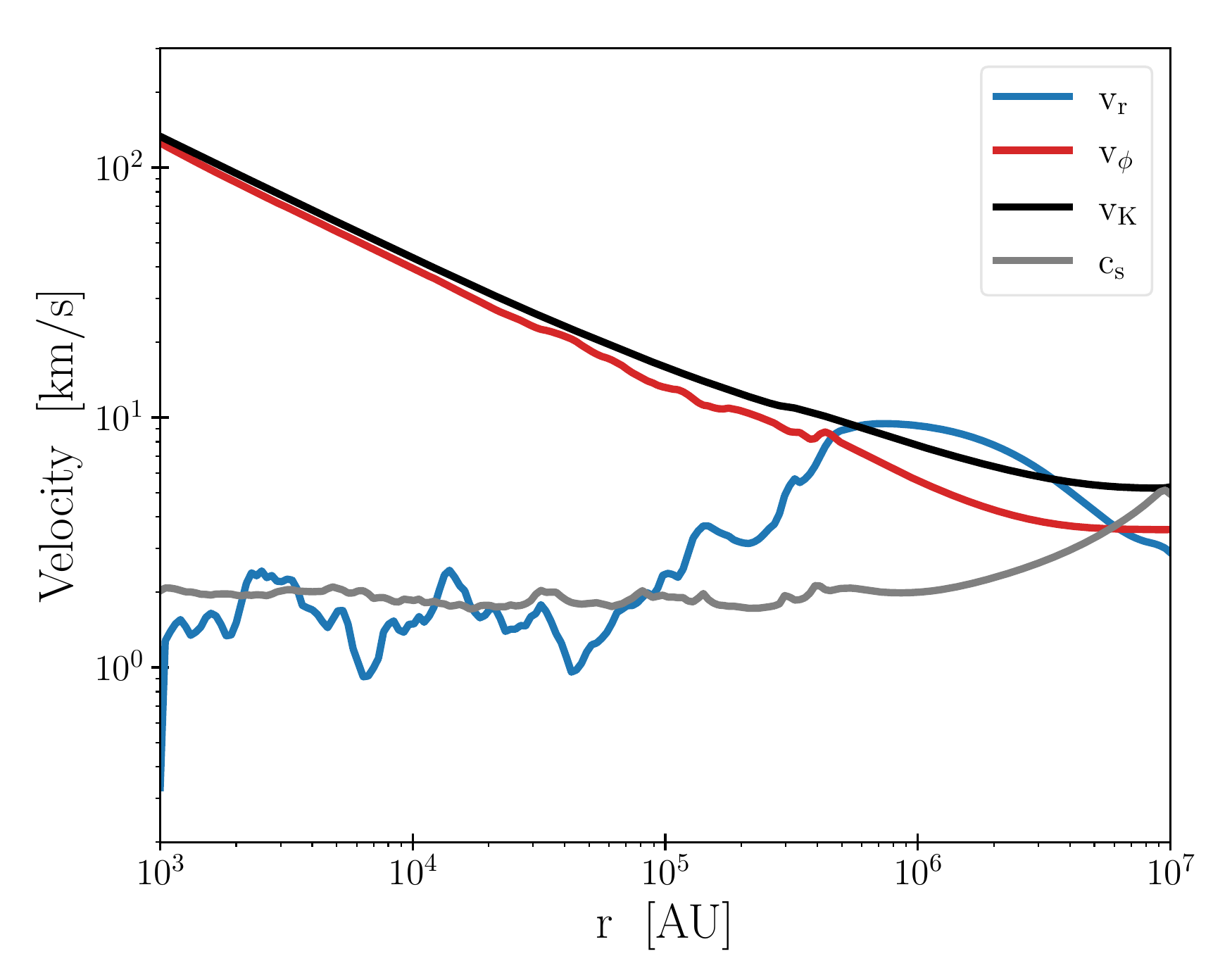}
\end{center}
\vspace{-5mm}
\caption{
Radial profiles of the time ($0\leq t/ \leq 1~\Myr$) and azimuthally ($0\leq \phi \leq 2\pi$) averaged velocity fields of the accretion flow along the equator 
(i.e., $\theta = \pi/2$) in the MMH halo: the radial velocity (blue), rotational velocity (red), Keplerian velocity (black), 
and sound speed (grey), respectively.
The time-averaged radial velocity shows the inflowing component at all radii (i.e., $v_r<0$).
}
\label{fig:vel_MMH}
\end{figure}

Figure~\ref{fig:vel_MMH} presents the time-averaged radial profiles of the rotational (red) and radial (blue) velocities along the equator.
For comparison, we overlay the Keplerian velocity (black) and the sound speed (grey).
Within the disk, the rotational velocity asymptotically approaches the Keplerian value, which indicates that the accretion disk is rotationally supported.
Note that the Kepler velocity follows $v_{\rm K} =(GM_\ast /r)^{1/2}$ in the inner region of the disk, where the gravity is dominated by the central star.
The time-averaged radial velocity has only the inflowing component (i.e. $v_r < 0$) and has a nearly constant value comparable to the sound speed at $r< 10^5~\au$.
Fast inflows with $v_r \lesssim c_{\rm s}$ are generally found in self-gravitating disks.
Based on the balance between the gravitational torque driven by bar/spirals and the advection of angular momenta along the accretion disk, 
the inward mass transport occurs at a few $10~\%$ of the sound speed in the disk \citep[][]{Goodman2003MNRAS}.

The distribution of the density and inflow velocity within the disk can be explained as follows.
Let us suppose that the density follows a single power-law profile of $\rho \propto r^{-n}$ and the sound speed of gas
is nearly constant in the disk, yielding $\Sigma \equiv 2 \rho H \propto r^{3/2-n}$.
In a quasi-steady state, mass conservation requires $v_r \propto r^{n-5/2}$, which self-consistently explains the density slope of $n = 5/2$ and the constant radial velocity in the simulation.
Furthermore, it is worth mentioning that the disk following the density slope would be most gravitationally unstable to perturbations with a characteristic wavenumber of $|k_{\rm c}| = \pi G \Sigma / c_{\rm s}^2 \propto r^{-1}$.
Then, if the angular momentum transfer along the disk is governed by turbulent motions associated with gravitational instability, the effective kinematic viscosity is expressed as $\nu \propto c_{\rm s}/|k_{\rm c}| \propto r$,
leading to a constant radial velocity as $v_r \propto \nu/r = {\rm const}$.
Thus, the nature of the circumstellar disk in the MMH halo can be characterized with transport of mass and angular momentum owing to gravitational instability.

\begin{figure}
\begin{center}
\includegraphics[width=\columnwidth]{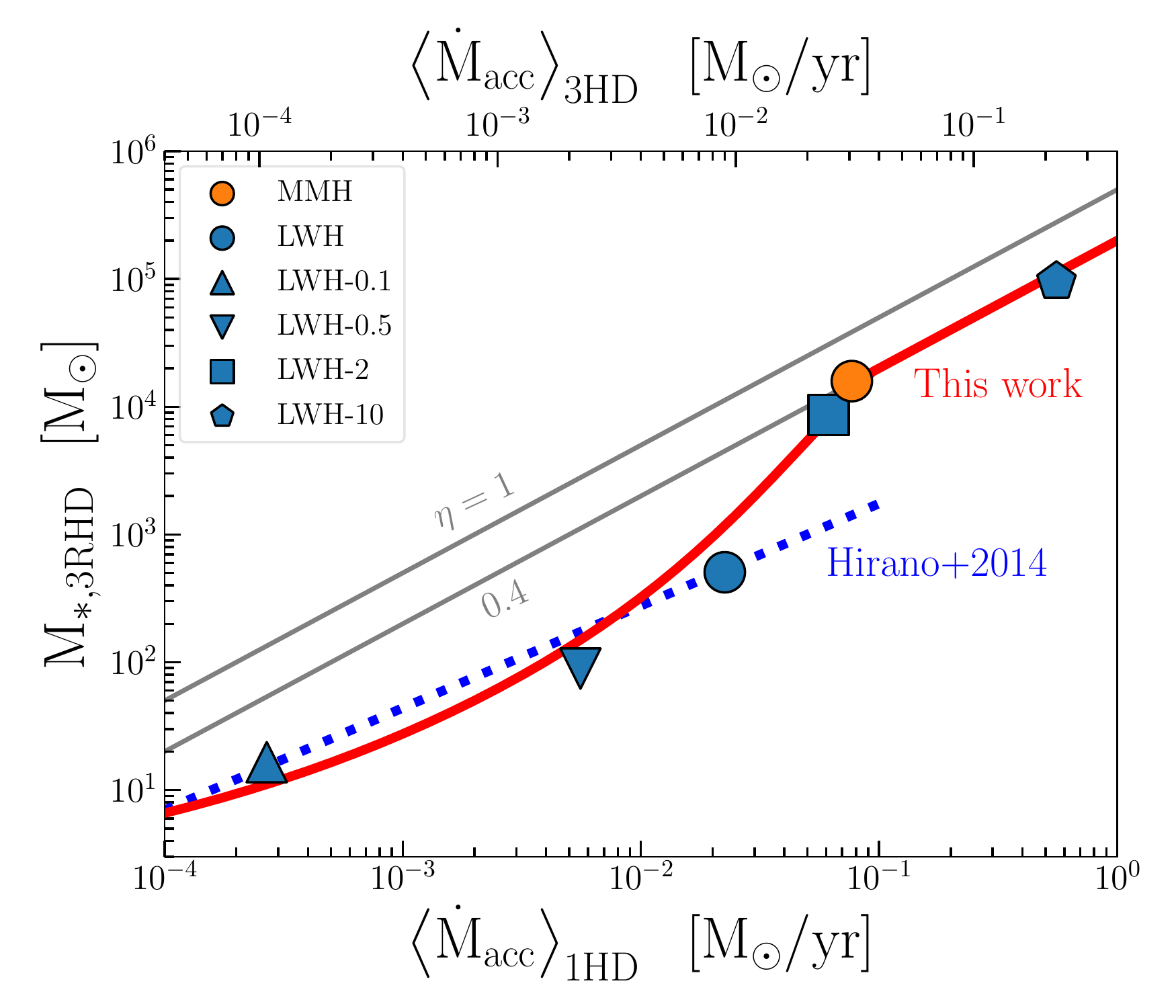}
\end{center}
\vspace{-5mm}
\caption{
Relation between the time-averaged mass accretion rate obtained in 1D HD simulations, $\langle \mdot \rangle_{\rm 1HD}$, 
and the mass of the central star in 3D RHD simulations, $M_{\ast, \rm 3RHD}$.
The values plotted here are evaluated at $t = 0.5~\Myr$, as shown in the second and fifth column of Table~\ref{table:model}.
The grey lines represent the mass growth efficiency of $\eta \equiv M_{\ast, \rm 3RHD}/M_{\ast, \rm 1HD} = 1$ and $0.4$, respectively. 
We also show $\langle \mdot \rangle_{\rm 3HD} \equiv 0.4~\langle \mdot \rangle_{\rm 1HD}$ on the upper $x$-axis, as a proxy for the mass accretion rate 
reduced by disk formation in the absence of radiative feedback.
The red solid curve presents the relation for the two quantities obtained from our analytical formula.
For comparison, we overlay the fitting relation obtained from the 2D RHD simulations of PopIII star formation \citep[blue dashed curve;][]{Hirano2014ApJ}.
}
\label{fig:ms_acc}
\end{figure}


\section{Mass growth of PopIII stars under various accretion histories}\label{sec:mass_growth}

In addition to the two cases discussed in \S~\ref{sec:LWH} and \ref{sec:MMH}, we explore four more variants in the LWH halo, 
where the initial density distribution is rescaled.
The simulation results for these cases are summarized in the last four lines of Table~\ref{table:model}.
In the LWH-2 and LWH-10 models (i.e., high density), the central star grows in mass efficiently and thus grow to be $\ms \simeq 9\times10^3~\msun$ and $4\times10^4~\msun$, respectively.
Those values are just a few times lower than those in 1D HD simulations, suggesting that stellar radiative feedback plays a minor role in determining the final stellar mass.
In contrast, in the LWH-0.1 and LWH-0.5 models (i.e., low density), further mass growth is prevented by radiative feedback above 
$\ms \simeq 20~\msun$ and $90~\msun$, respectively, while the protostar could increase its mass by a factor of $\gtrsim 10$ without radiative feedback.

Figure~\ref{fig:ms_acc} presents the relation between the time-averaged mass accretion rate found in the 1D HD simulation
and the final mass of the central accreting star calculated in the full 3D RHD simulation (see also Table~\ref{table:model}).
For reference, we define the accretion efficiency as $\eta \equiv M_{\ast, \rm 3RHD}/M_{\ast, \rm 1HD}$.
The value of $\eta$ characterizes the suppression level of mass accretion owing to angular momentum of the inflowing gas
and stellar radiative feedback.
For the low density cases with $\langle \mdot\rangle_{\rm 1HD} < \mdotc$, 
the resultant mass is substantially reduced by a factor of 20 (i.e., $\eta \sim 0.05$) from that in the 1D non-radiation case.
The suppression is mainly caused by radiative feedback from the contracting protostar with a high surface temperature of $T_{\rm eff}\simeq 10^5~{\rm K}$.
The simulation results (LWH-0.1, LWH-0.5, and LWH) are in good agreement with the fitting result (blue dashed) obtained by 
previous simulations of PopIII star formation in mini-haloes, where the mass inflow rate tends to be lower than $\mdotc$ \citep{Hirano2014ApJ}.
On the other hand, in the high density cases with $\langle \mdot\rangle_{\rm 1HD} > \mdotc$, 
the final mass increases with the inflow rate and approaches a constant efficiency with $\eta = 0.4$.
The jump of the efficiency occurs because such a rapidly accreting protostar evolves with a bloated envelope 
with $T_{\rm eff}\simeq 5000~{\rm K}$ and thus hardly emits ionizing radiation.
Note that $\eta = 0.4$ represents suppression of stellar mass growth by the finite angular momentum of the accretion flows 
(see also the upper grey line for $\eta = 1$ to highlight the effect of angular momentum).
In the following, we refer to $\langle \mdot \rangle_{\rm 3HD} \equiv 0.4~\langle \mdot \rangle_{\rm 1HD}$ as a mass accretion rate reduced from the spherical 1D case owing to disk formation but not to stellar radiative feedback,
shown in the upper horizontal axis of Figure~\ref{fig:ms_acc}.

Here, we interpret our simulation results with an analytical estimate of the mass loss rate 
from a circumstellar disk owing to photoevaporation \citep{Tanaka2013ApJ};
\begin{eqnarray}
\dot{M}_{\rm pe} \simeq 1.5 \times 10^{-2}~\msun~\yr^{-1}~
\left ( \frac{\dot{Q}_{\rm ion}}{10^{52}~{\rm s^{-1}}} \right)^{1/2}~
\left ( \frac{R_{\rm pe}}{10^{4}~{\rm AU}} \right)^{1/2} \ ,
\label{eq:pe_rate}
\end{eqnarray}
where $R_{\rm pe}$ is the physical scale within which the disk suffers from photoevaporation.
The value of $R_{\rm pe}$ depends on the size of the ionized region surrounding the star, disk size, and their geometrical configuration \citep{McKee2008ApJ, Hosokawa2011Sci}.
We here adopt $R_{\rm pe} = 10^4~{\rm AU}$ as a fiducial value, which is broadly consistent with the spatial extent of the photoevaporating region 
seen in the LWH, LWH-0.5, and LWH-0.1 models.
The production rates of ionizing photons, $\dot{Q}_{\rm ion}$, from a ZAMS star depend on stellar mass, as shown in Table~\ref{table:zams}.
Equating $\dot{M}_{\rm pe}$ with $\langle \mdot\rangle_{\rm 3HD}$, we obtain the equilibrium mass,
above which the stellar mass growth ceases because the circumstellar disk is evacuated by mass loss owing to photoevaporation.
For the high accretion-rate regime ($\langle \mdot\rangle_{\rm 3HD} > \mdotc$), the value of $\dot{Q}_{\rm ion}$ sharply drops and thus the mass loss process shuts off.
Therefore, the final stellar mass is not limited by radiative feedback but by the mass budget in the circumstellar disk or a short lifetime of the central massive star 
\citep[for comparison to 1D cases, see ][]{Omukai2003ApJ, Johnson2012ApJ}.
We find that the analytic model for disk photoevaporation (red curve) nicely explains the overall trend of our simulation results.

\begin{figure}
\begin{center}
\includegraphics[width=\columnwidth]{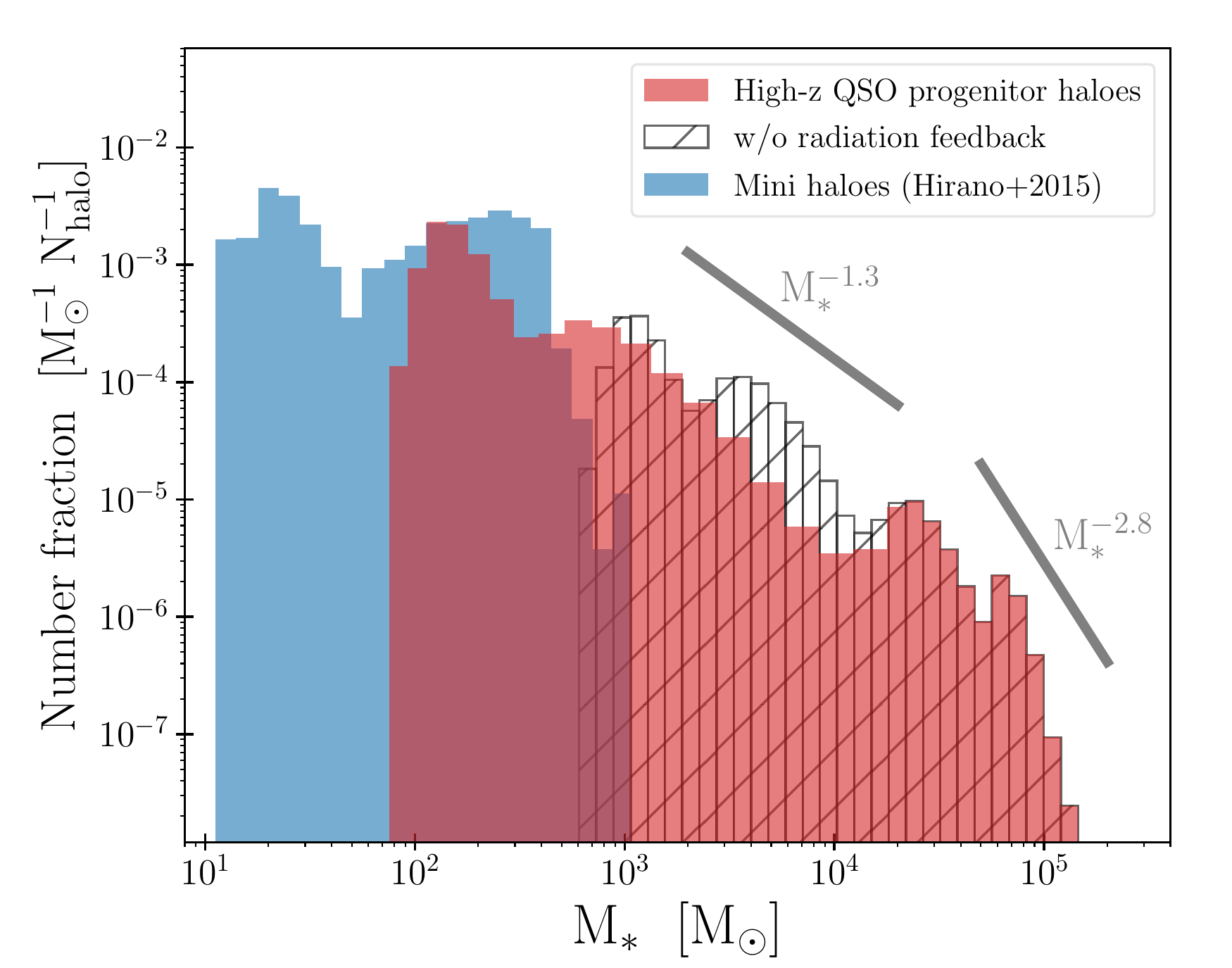}
\end{center}
\vspace{-5mm}
\caption{
The mass distribution function of massive primordial stars in a high-$z$ QSO progenitor halo (red histogram)
obtained from the $\langle \mdot \rangle_{\rm 1HD} - M_{\ast, \rm 3RHD}$ correlation (see Figure~\ref{fig:ms_acc}).
The probability distribution function of $\langle \mdot \rangle_{\rm 1HD}$ is taken from a merger-tree based 
semi-analytic model for BH seeding mechanisms \citep[][see the left-bottom panel of their Figure~6]{Li2021ApJ}.
The mass function can be approximated as $\propto \ms^{-1.3}$ at $\ms \leq 2 \times 10^4~\msun$ and $\propto \ms^{-2.8}$ at the higher mass, as indicated by the two grey lines.
To show the impact of radiative feedback, we overlay the mass distribution assuming a constant efficiency of $\eta =0.4$ 
in the limit of the absence of radiation (white hatched histogram). 
For comparison, the mass distribution of ordinary PopIII stars formed in a typical mini-halo is shown by the blue histogram
\citep{Hirano2015MNRAS}.
}
\label{fig:mf}
\end{figure}


\section{Discussion}\label{sec:discussion}

\subsection{Implication for the IMF of primordial stars}\label{sec:imf}

In this section, we discuss the initial mass function for primordial stars, applying the $\langle \mdot \rangle$ -- $M_\ast$ relation 
(the red curve in Figure~\ref{fig:ms_acc}) to star formation episodes in high-$z$ QSO host galaxies.
Recently, \citet{Li2021ApJ} have conducted a semi-analytic study of chemo-thermal dynamics of collapsing gas clouds,
using merger trees to trace the halo growth in overdense regions of the universe with $4\sigma$ overdensity.
They find that high-$z$ QSO progenitor haloes are likely irradiated by intense H$_2$-photodissociating radiation from nearby star-forming galaxies
and heat the interior gas by successive halo mergers.
As a result, the mass accretion rates of collapsing parent clouds show a great diversity,
depending on their evolutionary history and external environments (see Figure~8 in \citealt{Li2021ApJ}). 
Note that this accretion rate can be approximated as $\langle \mdot\rangle_{\rm 1HD}$ 
since their calculation does not take into account either ther angular momentum of the collapsing cloud or stellar radiative feedback.

Figure~\ref{fig:mf} shows the mass function of primordial stars formed in high-$z$ QSO host galaxies (red histogram).
The mass is widely distributed over $100~\msun < \ms < 10^5~\msun$, reflecting the diversity in the mass accretion rate. 
The shape of the mass distribution is approximated with a double power law function; 
$\propto \ms^{-1.3}$ at $\ms \leq 2 \times 10^4~\msun$ (significantly flatter than a Salpeter IMF of $\propto \ms^{-2.35}$ 
for metal-enriched stellar populations) and $\propto \ms^{-2.8}$ in the higher mass range.
We also present the mass function without including radiative feedback, but assuming a constant efficiency of $\eta =0.4$ 
(white hatched histogram).
This clearly demonstrates that radiative feedback creates the low-mass population below $\ms \lesssim 10^3~\msun$. 
Note that the number fraction of SMSs heavier than $\ms = 10^4~\msun$ is $\sim 20\%$, independent of the feedback process.
Since the total number density of DM haloes of interest in a comoving volume is estimated as $N_{\rm h}\sim 10^{-3}~{\rm cMpc}^{-3}$ \citep{Li2021ApJ},
the number density of SMSs (presumably, heavy BH seeds) is given by $N_{\rm SMS}\sim 2\times 10^{-4}~{\rm cMpc}^{-3}$.
Therefore, if $\sim 0.01\%$ of the SMSs collapse to BHs and grow in mass subsequently (see also \S~\ref{sec:BHs}), the number density of high-$z$ QSOs 
($N_{\rm SMBH}\sim 10^{-8}~{\rm cMpc}^{-3}$) can be explained.

For comparison, we also show the mass function of typical PopIII stars that form in DM haloes in less-biased regions of the universe,
where the external environmental effects (e.g., LW radiation and halo mergers) are modest (blue histogram taken from \citealt{Hirano2015MNRAS}).
This mass function occupies a relatively low-mass range with a mean value of $\langle \ms \rangle \simeq 220~\msun$, 
exhibiting a sharp cutoff at the high-mass end around $\ms \sim 10^3~\msun$.
This suggests the importance of external effects suppressing cloud collapse to extend the high-mass end of the mass distribution. 

We emphasize that normal PopIII stars are the representative population of primordial stars, while the heavier population
in rarer regions would play a more important role in the formation of SMBHs.
Numerical simulation studies of PopIII star formation provide a possible mass budget of PopIII stars,
$\rho_\ast \simeq 10^{5-6}~\msun~{\rm cMpc}^{-3}$, above which the emergent UV radiation would complete 
cosmic reionization earlier and yield a high optical depth of the universe to electron scattering inconsistent 
with the Planck measurement \citep{Visbal2020ApJ, Inayoshi2021ApJ}.
With the mean mass of PopIII stars in \citet{Hirano2015MNRAS}, the number density of PopIII stars in a cosmic volume is
as high as $N_{\rm PopIII} \simeq \rho_\ast/\langle \ms \rangle \sim 500-5,000~{\rm cMpc}^{-3}$.
Therefore, the formation efficiency of SMSs is limited to $\lesssim 10^{-7}$ of that of normal PopIII stars.

\subsection{Subsequent mass growth of remnant BHs}\label{sec:BHs}

Massive stars with $\ms \gtrsim 30~\msun$ are expected to leave BHs behind after the end of their lifetime, 
except in the mass range of $\ms \sim 140-260~\msun$, where pair-instability supernovae take place \citep[see][for a review]{Nomoto2013ARA&A}.
For metal-free primordial stars, which avoid significant mass loss by stellar winds, a large fraction of their initial mass ends up in the remnant BHs, 
i.e., $\mbh/\ms \sim 0.5$--$0.9$ \citep[e.g.,][]{Heger2003ApJ, Belczynski2010ApJ, Spera2015MNRAS}.
Therefore, both central stars formed in the LWH and MMH haloes will eventually leave remnant BHs 
with approximately the original stellar mass.

We here briefly discuss the subsequent gas accretion onto the stellar remnants in our simulated haloes.
In the LWH halo, a BH with $\mbh \sim 500~\msun$ would form at the center of the ionized cavity seen in Figure~\ref{fig:str_LWH}.
After the BH formation, since the central BH does not radiate by itself unlike a star, the hot ionized bubble cools down and collapses 
to the center without being impeded by radiative feedback.
In this case, the BH is fed by large-scale inflowing gas from $R \sim 10^6~\au$ at rates of $\dot{M} \gtrsim 0.05~\msun~\yr^{-1}$
as seen in Figure~\ref{fig:cm_LWH}.
Such rapid mass inflows would reach the central BH after a few free-fall timescales, i.e., $t > 1~\Myr$.
Although, in general, mass accretion rates are limited to the Eddington rate,
$\dot{M}_{\rm Edd} \simeq 10^{-5}~\msun~\yr^{-1}~\left( \mbh/500~\msun \right)$,
our previous simulations in \citet{Toyouchi2021ApJ} have demonstrated that a BH accretes gas at super-Eddington rates
overcoming radiative feedback associated with its accretion, when the mass supply rate from the circum-BH disk exceeds 
the critical value described as
\begin{eqnarray}
\dot{M}_{\rm \bullet, crit} \simeq 4.4 \times 10^{-2}~\msun~\yr^{-1}~\left( \frac{c_{\rm s}}{2~{\rm km~s^{-1}}} \right) \ ,
\label{eq:mc_for_edd}
\end{eqnarray}
where $c_{\rm s}$ is the sound speed in the disk.
We thus predict that after the central star turns into a BH in the LWH halo, the ionized cavity in the accretion disk is filled, 
and the remnant BH efficiently grows by rapid mass accretion.

In the case of the MMH halo, where an IMBH with $\mbh \sim 3\times10^4~\msun$ would form,
the mass accretion rate at $t \sim 1~\Myr$ is slightly lower than $\dot{M}_{\rm \bullet, crit}$.
Similar to the LWH halo, however, the mass supply rate onto the circum-BH disk would increase with time owing to 
mass inflows from larger radii.
Once the conditions for super-Eddington mass accretion are satisfied, the central BH becomes a milli-QSO radiating with 
$L_{\rm bol} \gtrsim L_{\rm Edd} \simeq 4 \times 10^{42}~{\rm erg~s^{-1}} (\mbh / 3 \times 10^4 ~\msun)$.
While the IMBH in the MMH halo located at $z \sim 15$ would be too faint to be observed, 
more luminous milli-QSOs with $L_{\rm bol} \sim 10^{45}~{\rm erg~s^{-1}}$, resulting from higher BH mass $\mbh \gtrsim 10^6~\msun$
and/or extremely high-Eddington ratios reach the detection limit for upcoming observations by the James Webb Space Telescope \citep[][]{Inayoshi2022barXiv}.
A future discovery of such milli-QSOs at $z > 10$ will be strong evidence of IMBH formation in the primordial universe and its rapid mass growth to establish SMBH populations found at $z \sim 7$.

\subsection{Effects of disk fragmentation and off-center star formation}\label{sec:unconsidered}

We here discuss the effects of disk fragmentation on the mass distribution of primordial stars.
While our grid configuration is adopted to resolve the gas dynamics at the vicinity of the central star (namely, the innermost radius),
disk fragmentation owing to gravitational instability potentially yields multiple stars surrounding the central star.
Such off-center star formation depletes gas in the circumstellar disk and makes the central star less massive than 
our prediction in Figure~\ref{fig:ms_acc} \citep[e.g.,][]{Inayoshi2014bMNRAS, Chon2018MNRAS}.
\citet{Sugimura2020ApJ} conducted 3D RHD simulations with an adaptive mesh refinement technique 
and found that formation of a twin binary (i.e., nearly equal mass) would likely take place under circumstances 
of ordinary PopIII star formation.
They reported that the total mass of the binary is reduced by a factor of $\sim 2$ from the central stellar mass
found in other 3D RHD simulations by \citet{Hosokawa2016ApJ} that adopt the same simulation setup and initial conditions 
except using spherical coordinates\footnote{
This result does not mean that disk fragmentation and off-center star formation cannot be calculated properly in a spherical coordinate grid.
Indeed, \citet{Oliva2020A&A} performed spherical coordinate based 3D RHD simulations with an extremely high spatial resolution 
and successfully demonstrated the formation of gravitationally collapsing cores in off-center regions of the accretion disk.}.
%
Therefore, this result implies that disk fragmentation would lead to a less top-heavy IMF, which is weighted towards 
$\ms < 100~\msun$ compared to that shown in Figure~\ref{fig:mf}.

However, in a massive cloud collapsing at a high rate of $\mdot > 0.1~\msun~{\rm yr}^{-1}$ onto the cloud center, 
the circumstellar disk is so dense that gas clumps formed by disk fragmentation can quickly migrate inward via gaseous dynamical friction.
Since the migration timescale is generally shorter than the KH timescale of the forming clumps, most clumps can plunge into 
the central protostar before forming massive companions \citep[][]{Inayoshi2014bMNRAS}.
Indeed, recent simulations also demonstrate that 
the most massive star located at the center is preferentially fed by clump migration and mergers with low-mass companions,
resulting in the formation of a single SMS \citep[e.g., ][]{Suazo2019ApJ, Chon2020MNRAS}.
Furthermore, \citet{Regan2020bOJAp} post-processed the impact of stellar radiation in their cosmological hydrodynamical simulations 
and found that ionizing radiation from survived companion stars is too weak to suppress the growth of the primary SMS.
Thus, we expect that our prediction of the PopIII IMF at $\ms \sim 10^4$--$10^5~\msun$ is not substantially affected by formation of low-mass companion stars 
associated with disk fragmentation.
To confirm this expectation, we will perform more sophisticated simulations that includes off-center star formation in the numerical domain in a future study.


\section{Summary}\label{sec:summary}

This paper presents a series of three-dimensional radiation hydrodynamical (3D RHD) simulations that study the formation processes of primordial supermassive stars (SMSs).
In particular, our simulations focus on the two pristine haloes (the LWH and MMH model), which were originally identified and studied by recent cosmological simulations of protogalaxy formation \citepalias[][]{Wise2019Nature}. 
These haloes are expected to be possible sites forming SMSs with masses of $\ms > 10^4~\msun$ since 
gravitational collapse of their gas clouds is suppressed by dynamical heating via rapid successive halo mergers until their virial masses reach $M_{\rm vir} \gtrsim 10^7~\msun$.
Unlike previous multi-dimensional simulations of SMS formation, 
our simulations successfully investigate the long-term stellar evolution up to $t = 1~\Myr$, almost the lifetime of a massive star,
following stellar evolution under variable mass accretion and radiative transfer of ionizing photons emitted from the growing star self-consistently.

For the LWH halo, we find that in the early phase ($t < 0.2~\Myr$), 
the central protostar is efficiently fed by dense accretion flows via spiral waves in the circumstellar disk
and grows to $\ms \sim 500~\msun$.
However, the mass accretion rate subsequently drops far below the critical value of $\mdotc = 0.04~\msun~\yr$,
below which the star contracts by losing its energy and evolves to a main sequence phase with a high effective temperature of $\teff \sim 10^5~\kelvin$. 
As a result, the massive central star continuously emits intense ionizing radiation and shuts down further mass growth, evacuating the circumstellar disk via photoevaporation.
Thus, the LWH halo fails to produce an SMS, in contrast to expectations from previous simulations \citepalias[e.g.,][]{Wise2019Nature, Sakurai2020MNRAS}.

On the other hand, the central star in the MMH halo accretes gas efficiently over the entire lifetime, eventually growing to an SMS with $\ms \sim 3\times10^4~\msun$.
In this case, since short accretion bursts exceeding $\mdotc \simeq 0.04~\msun~\yr^{-1}$ successively take place with intervals shorter than the Kelvin-Helmholtz timescale, 
the star can evolve in a cool bloating phase with $\teff \sim 5000~\kelvin$ without contraction.
As the bloated star hardly emits ionizing radiation, the impact of radiative feedback in the MMH halo is substantially weaker than in the LWH halo 
even though the central star becomes much heavier.

To cover a wide parameter space and understand the formation of massive primordial stars under various circumstances, 
we also examine four additional cases named LWH-0.1, -0.5, -2, and -10, for which the gas density in the LWH halo is rescaled by a factor of 0.1, 0.5, 2, and 10, respectively.
Compiling all the cases, we derive the relation between the final stellar mass $M_{\rm \ast, 3RHD}$ in the 3D RHD simulations and 
the time-averaged mass supply rate $\langle \dot{M}_{\ast} \rangle_{\rm 1HD}$ for the corresponding 1D cases 
in the absence of radiative feedback (see Figure~\ref{fig:ms_acc}).
This relation clearly highlight the suppression in stellar mass growth by radiative feedback and finite angular momentum of the accretion flows.
We find that the final stellar mass monotonically rises from $M_{\rm \ast, 3RHD} \sim 10~\msun$ to $10^5~\msun$ 
with increasing mass supply rates from $\langle \dot{M}_{\ast} \rangle_{\rm 1HD} \sim 10^{-4}~\msun~\yr^{-1}$ to $1~\msun~\yr^{-1}$.
In the LWH-0.1, LWH-0.5, and LWH models, where $\langle \dot{M}_{\ast} \rangle_{\rm 1HD} < \mdotc$, photoevaporation of the circumstellar disk effectively 
reduces the final stellar mass by a factor of a few tens compared to the mass expected in the absence of radiative feedback (Figure~\ref{fig:ms_acc}).
On the other hand, in the MMH, LWH-2, and LWH-10 models, where $\langle \dot{M}_{\ast} \rangle_{\rm 1HD} < \mdotc$, the final stellar mass is not 
determined by ionizing radiation feedback, but by the mass budget in the circumstellar disk or the short stellar lifetime.
It is also worth noting that for the more rapid accretion cases, the stellar mass obtained in the 3D simulations is reduced by a factor of 0.4 compared to that in the 1D ones, which represents suppression of stellar mass growth by centrifugal forces exerted on the accretion flows.

Finally, we discuss the initial mass function of primordial stars formed in massive haloes associated with rare density peaks that end up in quasar host galaxies at $z \sim 7$.
Combining our simulation results with the statistical properties of parent clouds predicted by a semi-analytical model \citep{Li2021ApJ}, we find that the stellar mass is widely distributed over
$\ms = 100$--$10^5~\msun$ and the function shape is approximately given by a double power law with two slopes of $\propto \ms^{-1.3}$ and $\ms^{-2.8}$ across a characteristic mass of 
$\ms \sim 2 \times 10^4~\msun$.
The global average number density of SMSs expected from this mass function is estimated as $\sim 2\times 10^{-4}~{\rm cMpc}^{-3}$.
This implies that if as few as $\sim 0.01 \%$ of those SMSs collapse to intermediate-mass BHs and grow via efficient gas accretion, the observed number of SMBHs at $z \sim 7$ can be explained.
Based on the final state of the MMH model, the remnant BH is expected to be fed via the dense debris disk at super-Eddington accretion rates.


\section*{Acknowledgements}

We thank K.~Omukai and T.~Hosokawa for fruitful discussions, and 
K.~Sugimura, R.~Nakatani, and Y.~Sakurai for their contribution to developing the numerical code. 
The numerical simulations were performed with the Cray XC50 at the Center for Computational Astrophysics (CfCA) of the National Astronomical Observatory of Japan.
This work is supported in part by 
JSPS KAKENHI Grant Numbers 21K20378, 
the National Natural Science Foundation of China (12073003, 12003003, 11721303, 11991052, 11950410493, and 1215041030), 
the China Manned Space Project Nos. CMS-CSST-2021-A04 and CMS-CSST-2021-A06, 
NSF grant AST-2006176, 
the Heisenberg Research Grant funded 
No.~KU 2849/9, 
and the JSPS Invitational Fellowship for Research in Japan ID S20156.


\section*{Data availability}

The data underlying this article will be shared on reasonable request to the corresponding author.




\bibliographystyle{mnras}
\bibliography{refs.bib} 



\appendix



\section{Numerical checks}\label{sec:cc}

We present the results of three additional runs with the same initial condition as the LWH halo but different grid configurations.
In the LWH-HR run, we double the number of grid cells in the polar and azimuthal directions, i.e., ($N_r$, $N_\theta$, $N_\phi$) = (200, 72, 144)
to better capture the nature of disk fragmentation \citep[e.g.,][]{Federrath2011ApJ, Turk2012ApJ, Meece2014ApJ}.
In the LWH-HS run, we set a smaller sink cell with a size of $\rmin = 500~\au$ to see the effect of dense clouds that might form at the innermost radius
and attenuate the incident stellar radiation, resulting in suppression of disk photoevaporation \citep[e.g.,][]{Jaura2022MNRAS}.
Finally, in the LWH-NE run, we relax the assumption of equatorial symmetry since asymmetric vertical flows passing through the equator perturb 
the circumstellar disk and regulate the mass accretion onto the central star \citep[e.g.,][]{Toyouchi2021ApJ}.

\begin{figure}
\begin{center}
\includegraphics[width=\columnwidth]{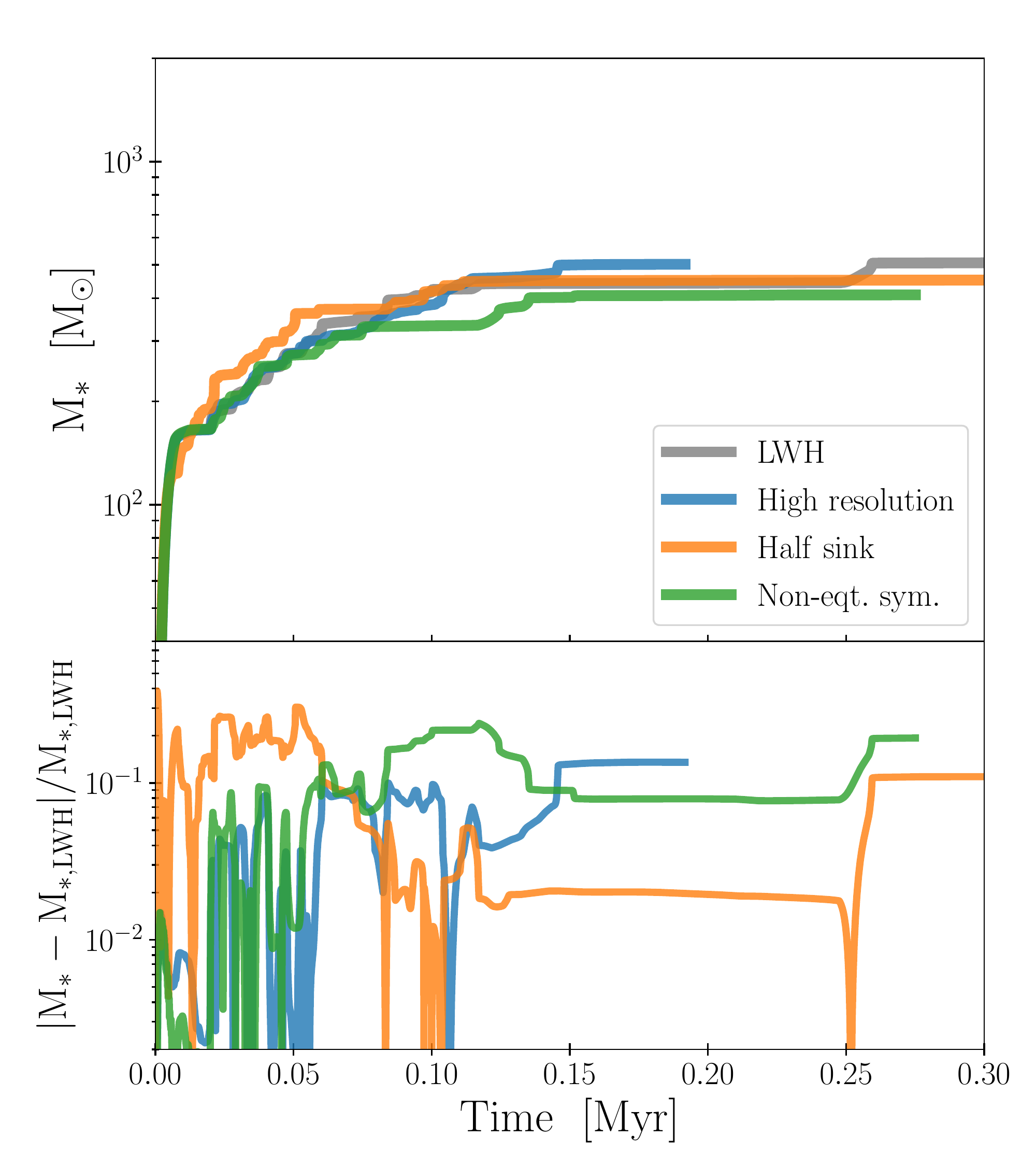}
\end{center}
\caption{
{\it Top panel}: 
Mass growth histories of the central star obtained in the LWH-HR (blue), LWH-HS (orange), and LWH-NE models (green), which adopts twice the number of cells in the tangential directions, halves the sink radius, and relaxes the assumption of equatorial symmetry, respectively, compared to the original simulation for the LWH halo (grey).
{\it Bottom panel}: 
The corresponding mass deviation in each case from the LWH model as a function of time.
We find that the stellar mass growth history is almost identical among these calculations with less than 20~\% difference at least until $t \sim 0.2~\Myr$.
}
\label{fig:acc_hr}
\end{figure}

In Figure~\ref{fig:acc_hr}, we present the time evolution of the stellar mass for the three models as well as the original LWH model.
In all of these cases, the stellar mass increases up to $\ms \sim 400~\msun$ in the early stage, but the mass growth halts around $t = 0.15~\Myr$.
The overall trend is almost identical to the original result in the LWH model.
Thus, our simulation result in the LWH case is not sensitive to the numerical treatments discussed above.

It is worth speculating whether the stellar mass growth history changes with further increasing spatial resolution and reducing sink radii.
According to 3D RHD simulations by \citet{Oliva2020A&A} that studied disk fragmentation and star formation with an extensive convergence test varying the grid resolution by $2^5$,
the evolution of the central stellar mass and the number of fragments quantitatively converge above their intermediate resolution level with $(N_{r},~N_\theta,~N_\phi) = (134,~41,~128)$, 
which is slightly coarser than that adopted in our LWH-HR model.
Thus, we conclude that our result is converged sufficiently.
On the other hand, 
with a minimal sink radius nearly reaching the surface of the central star, the outer disk regions might be shielded from stellar radiation by 
dense clumps that could form at the inner disk, resulting in reduction of mass loss owing to photoevaporation.
This issue is left for future investigation.

\bsp	
\label{lastpage}
\end{document}